\documentclass[12pt,preprint]{aastex}
\slugcomment{Accepted for Publication in the Astrophysical Journal}
\shorttitle{High-Energy Neutrinos in Shocked Pulsar Winds}
\shortauthors{S. Nagataki}

\begin{document}

\title{High-Energy Neutrinos Produced by Interactions of Relativistic Protons in Shocked Pulsar Winds}
\author{Shigehiro Nagataki\altaffilmark{1,2}}

\affil{$^1$Department of Physics, School of Science, University
of Tokyo, \\7-3-1 Hongo, Bunkyoku, Tokyo 113-0033, Japan}
\affil{$^2$ Research Center for the Early Universe, School of Science,
University of Tokyo,\\
7-3-1 Hongo, Bunkyoku, Tokyo 113-0033, Japan}
\email{nagataki@utap.phys.s.u-tokyo.ac.jp}

%%%%%%%%%%%%%%%%%%%%%%%%%%%%%%%%%%%%%%%%%%%%%%%%%%%%%%%%%%%%%%%%%%%%%%%%%%%%%%
\begin{abstract}
We have estimated fluxes of neutrinos and gamma-rays that are generated from decays of
charged and neutral pions from a
pulsar surrounded by supernova ejecta in our galaxy, including
an effect that has not been taken into consideration, that is, interactions
between high energy cosmic rays themselves in the nebula flow, assuming that
hadronic components are the energetically dominant species in the pulsar wind.
Bulk flow is assumed to be randomized by passing through the termination shock
and energy distribution functions of protons and electrons behind the
termination shock are assumed to obey the relativistic Maxwellians.
We have found that fluxes of neutrinos and gamma-rays depend very sensitively
on the wind luminosity, which is assumed to be comparable to the spin-down
luminosity. In the case where $B=10^{12}$G and $P=1$ms, neutrinos should be
detected by km$^3$ high-energy neutrino detectors such as AMANDA and IceCube.
Also, gamma-rays should be detected by Cherenkov telescopes such as CANGAROO
and H.E.S.S. as well as by gamma-ray satellites such as GLAST.
On the other hand,
in the case where $B=10^{12}$G and $P=5$ms, fluxes of neutrinos and
gamma-rays will be too low to be detected even by the next-generation
detectors. However, even in the case where $B=10^{12}$G and $P=5$ms, there
is a possibility that very high fluxes of neutrinos may be
realized at early stage of a supernova explosion ($t \le 1$yr), where the
location of the termination shock is very near to the pulsar.
We also found that there is a possibility that protons with energies $\sim 10^5$ GeV
in the nebula flow may interact with the photon field from surface of the
pulsar and produce much
pions, which enhances the intensity of resulting neutrinos and gamma-rays.
\end{abstract}
%%%%%%%%%%%%%%%%%%%%%%%%%%%%%%%%%%%%%%%%%%%%%%%%%%%%%%%%%%%%%%%%%%%%%%%%%%%%%

%%%%%%%%%%%%%%%%%%%%%%%%%%%%%
\section{Introduction}
%%%%%%%%%%%%%%%%%%%%%%%%%%%%%

It has been about 35 years since Goldreich and Julian (1969) pointed out
that a rotating magnetic neutron star generates huge electric potential
differences between different parts of its surface and, as a result,
should be surrounded with charged plasma, which is called a magnetosphere.
Gunn and Ostriker (1969) also pointed out the possibility that a rotating
magnetic neutron star may be a source of high energy cosmic rays.
Such high energy cosmic rays are considered to be driven along magnetic field
lines since these lines do not cross. Also, since part of the magnetic fields around
the neutron star passes through the light cylinder, which means such magnetic
fields are open, accelerated charged particles are driven to outside of the
light cylinder, which are called as pulsar winds. The pulsar winds are
usually considered to be composed of electron-positron pairs since electron-positron
pairs will be created so as to eliminate electric fields that are parallel to magnetic fields
at a region where net charge density is not equal to the Goldreich-Julian
density~\citep{ruderman75,shibata91}.
However, it is also pointed out that hadronic component may exist in pulsar
winds as a consequence of the net charge neutrality in the
outflow~\citep{hoshino92,bednarek97,protheroe98,blasi00,amato03}.
Moreover, it is pointed out that hadronic components may be the energetically dominant species
although they are dominated by electron-positron pairs in number~\cite{hoshino92}.
This is because inertial masses of hadrons are much larger than that of electron.

Based on the assumption that hadronic components are not negligible in pulsar winds,
some scenarios are proposed to produce high energy neutrinos and gamma-rays from
decays of charged and neutral pions that are produced by
interactions between hadronic, accelerated high energy cosmic rays and surrounding
photon fields and/or matter. Atoyan and Aharonian (1996) estimated flux of gamma-rays
and discussed its contribution on the observed spectrum of the Crab nebula,
although they concluded that its contribution may be important only at energies above
10 TeV. Bednarek and Protheroe (1997) proposed that accelerated
heavy nuclei can be photo-disintegrated in the pulsar's outer gap, injecting
energetic neutrons which decay into protons. The protons from neutron decay inside the
supernova ejecta should accumulate, producing neutrinos and gamma-rays in collisions with the
matter in the supernova ejecta.\footnote{They call this region as a nebula, which is slightly
different from the definition of a nebula given by Kennel and Coroniti (1984).
In this study, as explained below, we adopt the definition presented by Kennel and
Coroniti (1984).} There are some papers based on this scenario and flux of neutrinos
and/or gamma-rays is estimated~\citep{protheroe98,bednarek03a,bednarek03b,bednarek03c,amato03}.
Beal and Bednarek (2002) proposed that accelerated cosmic rays will interact with
the photon fields inside the supernova remnant at the very early phase (within $\sim$1 yr)
of a supernova explosion. Bednarek (2001) calculated the extragalactic neutrino background
based on this scenario.

In this study, we estimate fluxes of neutrinos and gamma-rays including
an effect that has not been taken into consideration, that is, interactions
between high energy cosmic rays themselves. This picture is based on the works
given by Rees and Gunn (1974) and Kennel and Coroniti (1984). Rees and Gunn (1974)
pointed out that the supersonic pulsar wind would terminate in a standing reverse
shock located at a distance $r_s$ from the pulsar. Beyond the termination shock,
the highly relativistic, supersonic flow is randomized and bulk speed becomes subsonic,
obeying the Rankine-Hugoniot relations. According to Hoshino et al. (1992), who studied
the theoretical properties of relativistic, transverse, magnetosonic collisionless shock
waves in electron-positron-heavy ion plasmas, proton
distribution functions in the down stream are found to be almost exactly described
by relativistic Maxwellians with temperatures $T_{p,2}/\gamma_1 m_p c^2 \sim 0.34$,
where $T_{p,2}$ is temperature at the down stream, $\gamma_1$ is the bulk Lorenz factor
of protons in the up stream. It is noted that protons are not thermalized through the
interactions with protons themselves, and/or electrons but just obeys the Maxwellian
distribution through transferring cyclotron waves. This subsonic flow speed would be, by
communicating with the nebula boundary at $r_N$ via sound waves, adjusted to match the
expansion speed of the supernova remnant (that is, supernova ejecta) at the innermost region.
This subsonic flow
is called as nebula flow in the study of Kennel and Coroniti (1984). We also adopt
this definition in this study.
%It is noted that synchrotron continuum of electrons
%is calculated and compared with the observation of Crab nebula by Kennel and Coroniti (1984b).
%Also, Atoyan and Aharonian (1996) calculated fluxes of synchrotron continuum and inverse
%compton spectrum based on the study of Kennel and Coroniti (1984).
In this study, we calculate flux of neutrinos and gamma-rays from charged and neutral pion decays
in the nebula flow which are produced through the interactions between high energy protons
themselves, assuming that energy distribution functions of protons obey the relativistic Maxwellians.

In this study, as the
previous works, we assume that protons are energetically dominant in the pulsar winds. 
Thus, we describe the nebula flow using the proton mass as a unit of mass.
We estimate pion production rates due to the interactions between high energy protons
using proper Lorenz transformations. As for the cross sections between protons, scaling
model~\cite{badhwar77} is adopted. Isobar model is not included in this study
since we consider production rates of high energy pions. Calculating the spectrum of neutrinos
and gamma-rays in the termination shock rest frame, we estimate these fluxes at the earth
assuming that the pulsar is located at 10 kpc away from the earth. 
%Number density of protons
%is assumed to be equal to that of electrons due to charge neutrarity. Initial bulk Lorenz
%factor of protons is assumed to be same with that of electrons. Thus, protons are energetically
%dominant since inertial mass of a proton is about 2$\times 10^3$ times larger than that of
%electron. 

In section~\ref{form}, we explain the formulation in this study. Results are
shown in section~\ref{results}. Discussions are given in section~\ref{discussions}.
Summary and conclusion are presented in section~\ref{conclusion}.

%%%%%%%%%%%%%%%%%%%%%%%%%%%%%%%%%%%%%
\section{FORMULATION}\label{form}
%%%%%%%%%%%%%%%%%%%%%%%%%%%%%%%%%%%%%

%---------------------------------------------------------------------------
\subsection{Confinement of Pulsar Wind by the Surrounding Supernova Remnant}
\label{flow}
%----------------------------------------------------------------------------
In this subsection, we briefly review the steady state, spherically symmetric,
magnetohydrodynamic model of nebula flow presented by Kennel and
Coroniti (1984), in which a highly relativistic pulsar wind is terminated 
by a strong MHD shock that decelerates the flow and increases its pressure
to match the boundary conditions imposed by the surrounding supernova 
remnant. In this study, we consider the possibility that the protons
are the energetically dominant species in the flow as pointed by Hoshino et
al. (1992). According to the work presented by Hoshino et al. (1992),
the proton distribution functions behind the termination shock are
almost exactly described by relativistic Maxwellians. Thus we can
define temperature behind the shock wave and solve the following nebula
flow using the magnetohydrodynamic equations.

%-----------------------------------------------------------------------
\subsubsection{Super-relativistic Magnetohydrodynamic Shock}\label{shock}
%-----------------------------------------------------------------------
It is usually introduced the parameter $\sigma$, which is the ratio of the
magnetic plus electric flux to the particle energy flux at just ahead of the
termination shock,
\begin{eqnarray}
\sigma = \frac{B^2}{4 \pi n u \Gamma m_p c^2},
\label{eqn1}
\end{eqnarray}
where $B$ is the amplitude of the magnetic field in the observer's frame, $n$ is the proper density,
$u$ is the radial four speed of the flow, $\Gamma$ is the bulk Lorenz factor
of the flow ($\Gamma^2 = 1 + u^2$), $m_p$ is the proton mass, $c$ is the
speed of light. It is assumed that the energy density of electric filed
is nearly equal to that of magnetic field. The luminosity $L$ at just
ahead of the termination shock is described as
\begin{eqnarray}
L = 4 \pi n \Gamma u r_s^2 m_p c^3 (1 + \sigma),
\label{eqn2}
\end{eqnarray}
where $r_s$ is the radius of the termination shock.
The termination shock is assumed to be stationary relative to the pulsar
and the observer. Thus the Rankine-Hugoniot relations for $90^\circ$ shocks
can be used directly to obtain the physical quantum in the observer's frame
behind the shock wave:
\begin{eqnarray}
n_1 u_1 = n_2 u_2
\label{eqn3}
\end{eqnarray}
\begin{eqnarray}
E = \frac{u_1B_1}{\Gamma_1} = \frac{u_2B_2}{\Gamma_2}
\label{eqn4}
\end{eqnarray}
\begin{eqnarray}
\Gamma_1  \mu_1 + \frac{EB_1}{4 \pi n_1 u_1} = \Gamma_2  \mu_2 + \frac{EB_2}{4 \pi n_2 u_2} 
\label{eqn5}
\end{eqnarray}
\begin{eqnarray}
\mu_1 u_1 + \frac{P_1}{n_1 u_1} + \frac{B_1^2}{8 \pi n_1 u_1} =
\mu_2 u_2 + \frac{P_2}{n_2 u_2} + \frac{B_2^2}{8 \pi n_2 u_2}.
\label{eqn6}
\end{eqnarray}
Subscripts 1 and 2 label upstream and downstream parameters. 
$E$ denotes the shock frame electric fields and $\mu = (e+P)/n$ is the specific enthalpy,
which for a gas with an adiabatic index $\gamma$ is defined
by
\begin{eqnarray}
\mu = mc^2 + \frac{\gamma}{\gamma -1} \left( \frac{P}{n}   \right).
\label{eqn7}
\end{eqnarray}
In this study, we set $\gamma_2$ to be 4/3, as explained in
Appendix~\ref{appa}. 

Kennel and Coroniti (1984) introduced a parameter $Y$ to obtain the
downstream radial four speed, $u_2$. Definition of $Y$ is
\begin{eqnarray}
Y \equiv \frac{B_2}{B_1} = \frac{\Gamma_2 u_1}{\Gamma_1 u_2}.
\label{eqn8}
\end{eqnarray}
Using this parameter and $\gamma_2 = 4/3$, Eq.~(\ref{eqn6}) can be expressed as
\begin{eqnarray}
\nonumber
&Y^2& - Y \left[   \frac{2}{\Gamma_2 u_2} \left( u_2^2 + \frac{1}{4}  
 \right) \frac{u_1}{\Gamma_1}   \right]
+
\left[   \frac{2}{\Gamma_2 u_2} \left( u_2^2 + \frac{1}{4}  
 \right) \left(  \frac{4 \pi n_1 \mu_1 \Gamma_1^2}{B_1^2}  
  \frac{u_1}{\Gamma_1} \right)  \right] \\
&& - \frac{2 \pi n_1 m_p c^2}{B_1^2} \frac{u_1}{u_2}  
 - \left( 1 + \frac{8 \pi (n_1 \mu_1 u_1^2 + P_1)}{B_1^2}  \right) = 0.
\label{eqn9}
\end{eqnarray}
Although there are some trivial typos for the expression of Eq.~(\ref{eqn9})
in the original paper, they does not affect the following discussion.

Assuming that the bulk Lorenz factor of the pulsar wind is sufficiently
large and the flow is sufficiently cold, we can approximate $\Gamma_1$,
$P_1$, and $\mu_1$ as $\Gamma_1 \sim u_1$, $P_1 \sim 0$, and $\mu_1
\sim m_p c^2$. Using these approximations, we can rewrite Eq.~(\ref{eqn9}) as
\begin{eqnarray}
u_2^2\left( u_2^2 + \frac{1}{4}   \right)^2 = (1 + u_2^2) \left( 
u_2^2 - \frac{1}{4}\frac{\sigma}{1+ \sigma}
\right)^2.
\label{eqn10}
\end{eqnarray}
Solving Eq.~(\ref{eqn10}), we can obtain the radial four speed behind the
shock wave as
\begin{eqnarray}
%\nonumber
u_2^2 = \frac{8\sigma^2 + 10 \sigma + 1}{16(\sigma + 1)} 
+ \frac{1}{16(\sigma + 1)}
\left[  64\sigma^2(\sigma+1)^2  + 20 \sigma (\sigma+1)+1    \right]^{1/2}.
\label{eqn11}
\end{eqnarray}
Note that the radial four velocity behind the shock wave is determined
by only one parameter, $\sigma$.

As for the pressure $P_2$, it can be obtained as
\begin{eqnarray}
\frac{P_2}{n_1m_pc^2u_1^2} = \frac{1}{4u_2\Gamma_2} \left[ 
1+ \sigma \left(  1 - \frac{\Gamma_2}{u_2} \right)
\right],
\label{eqn12}
\end{eqnarray}
where we assumed that $\mu_2 \sim 4(P_2/n_2)$.

In this study, we adopt a slightly different way to determine the
temperature from the way presented by Kennel and Coroniti (1984), using
the results of Hoshino et al. (1992). According to Hoshino et al. (1992),
the distribution functions of protons
behind the termination shock are almost exactly described by relativistic
Maxwellians as
\begin{eqnarray}
N(\gamma) = A \gamma \exp \left[   -\frac{m_pc^2}{k_B T} (\gamma -1)   \right],
\label{eqn13}
\end{eqnarray}
where $\gamma$ is the Lorenz factor that represents the random motion (note that
this is not the adiabatic index),
$k_B$ is the Boltzmann constant and $A$ is the normalization factor
in units of cm$^{-3}$. In this case, the relation 
\begin{eqnarray}
P_2 = \frac{2}{3} n_2 k_B T_2
\label{eqn14}
\end{eqnarray}
holds exactly (see Appendix~\ref{appa} in detail). Thus the temperature $T_2$
can be derived from Eqs.~(\ref{eqn12}) and~(\ref{eqn14}) as
\begin{eqnarray}
\frac{k_BT_2}{u_1m_pc^2} = \frac{3}{8 \Gamma_2} \left[ 
1+ \sigma \left(  1 - \frac{\Gamma_2}{u_2} \right)
\right].
\label{eqn15}
\end{eqnarray}

As a result, we can estimate the energy density ($e$) [erg cm$^{-3}$] as  
\begin{eqnarray}
e = Am_p c^2 e^{\alpha} \int_1^{\infty}   \gamma^2 e^{-\alpha \gamma}
d \gamma, 
\label{eqn16}
\end{eqnarray} 
where $\alpha = m_pc^2/k_BT$. Since we can obtain the physical quantum
just behind the termination shock, we consider the resulting flow from the
termination shock to the remnant. This flow is usually called as nebula flow.

%-----------------------------------------------------------------------
\subsubsection{Nebula Flow}\label{nebula}
%-----------------------------------------------------------------------
The steady and spherically symmetric nebula flow is described by the
equations mentioned below. These are;
the equation describing the conservation of number flux,
\begin{eqnarray}
\frac{d}{dr} (cnur^2) = 0;
\label{eqn17}
\end{eqnarray}
one describing the conservation of magnetic flux in the magnetohydrodynamic
approximation,
\begin{eqnarray}
\frac{d}{dr} \left(   \frac{ruB}{\Gamma}     \right) = 0;
\label{eqn18}
\end{eqnarray}
one describing the propagation of thermal energy,
\begin{eqnarray}
\frac{d}{dr} \left(  ur^2 e   \right) + P\frac{d}{dr} (r^2u) = 0  ;
\label{eqn19}
\end{eqnarray}
and one describing the conservation of total energy,
\begin{eqnarray}
\frac{d}{dr} \left[ nur^2 \left(   \Gamma \mu + \frac{B^2}{4 \pi n \Gamma}
    \right)   \right]  = 0.
\label{eqn20}
\end{eqnarray}
Note that the notation for
the thermal energy density is slightly different from the original paper.

When the relation $e = 3P$ holds, Eq.(\ref{eqn19}) reduces to
\begin{eqnarray}
\frac{d}{dr} \ln \left(   \frac{P}{n^{4/3}}    \right) = 0.
\label{eqn21}
\end{eqnarray}
These are the basic equations describing the nebula flow which connects
the condition behind the termination shock and the inner-edge of the supernova
remnant.

Here we assumed that the distribution functions of protons remain Maxwellian
and temperature can be determined at each radius of the nebula flow. However,
we should adopt the Boltzmann equations coupled with the evolution of magnetic
fields to obtain the exact solution for the nebula flow, since it is not
apparent for protons to transfer their energies with each other
enough to obey the Maxwellian distribution.
However, as shown in section~\ref{results}, charged and neutral
pions are mainly produced just behind the termination shock (see also
figures~\ref{fig8} and~\ref{fig12}), where
the energy distribution of protons obey relativistic Maxwellian.
The nebula flow can be described approximately as a free-expansion and
temperature and density in the nebula flow becomes smaller along with radius,
which means pions are less produced along with radius. Thus we consider that
the order-estimate of the neutrino flux produced by $pp$ collisions will be
valid even if the MHD equations are adopted.
We will estimate the neutrino flux using the Boltzmann equations coupled with
the evolution of magnetic fields in the forthcoming paper.

The solution for the nebula flow can be obtained analytically (see Appendix~\ref{appc} for
details). Using the solution, total pressure $P_{\rm T}$ in the postshock can be
expressed as
\begin{eqnarray}
P_{\rm T} = P + \frac{E^2+B^2}{8 \pi} = \frac{L}{4 \pi r_s^2 c (1+ \sigma)}
\left[  \frac{P_2}{n_1m_pc^2u_1^2}(vz^2)^{-4/3} + \frac{\sigma}{z^2}
\left(   1 + \frac{1}{2 u_2^2v^2}   \right)      \right],
\label{eqnadd6}
\end{eqnarray}
where $v$ is defined as $v = u/u_2$ and $z$ is defined as $z=r/r_s$.

As explained in section~\ref{boundary}, we investigate the case $\sigma \ll 1$,
because the speed of the nebula flow at the boundary between the remnant
and nebula flow is set to be almost same with the speed of the remnant
($\sim 2000$ km s$^{-1}$). In the case of $\sigma \ll 1$,
\begin{eqnarray}
u_2 \sim \left( \frac{1 + 9 \sigma}{8}   \right)^{1/2}
\label{eqnadd7}
\end{eqnarray}
and
\begin{eqnarray}
\frac{P_2}{n_1m_pc^2u_1^2} \sim \frac{2}{3}(1-7\sigma).
\label{eqnadd8}
\end{eqnarray}
Thus the total pressure can be written approximately
\begin{eqnarray}
P_{\rm T} \sim \frac{2}{3}(n_1\Gamma_1u_1m_pc^2) 
\left[ \frac{1-7\sigma}{(vz^2)^{4/3}} + \frac{\sigma}{z^2} \left(  1 + \frac{4}{v^2}
   \right)     \right].
\label{eqnadd9}
\end{eqnarray}
This approximation is used in section~\ref{boundary}.

%--------------------------------------------------
\subsubsection{Boundary Conditions}\label{boundary}
%--------------------------------------------------

As for the inner boundary condition, the parameters are bulk Lorenz factor
and luminosity of the wind. As for the bulk Lorenz factor, we can estimate
its upper limit as a function of the rotation period and the amplitude of the polar
magnetic field of the pulsar~\cite{goldreich69}. The corotating magnetosphere
is bounded by a field line whose feet are at $\sin \theta_\circ
\sim  (\Omega R/c)^{1/2} = (2R/P)^{1/2} \times 10^{-5}$, where
$\theta_\circ$ is the zenith angle (in a unit of rad), $\Omega$ is the angular velocity of the
pulsar, $R$ is the radius (in a unit of cm), and $P$ is the rotation period (in a unit of s). The potential
difference between $\theta_\circ$ and the pole is
\begin{eqnarray}
\Delta \Phi = \frac{1}{2} \left(   \frac{\Omega R}{c}    \right)^2 R B_p
\label{eqn22}
\end{eqnarray} 
for $\Omega R/c \ll 1$ when the surrounding region of the pulsar is vacuum.
Here $B_p$ is the amplitude of the polar magnetic field in units of G. 
Although the magnetosphere should be filled with plasma whose number density
is the Goldreich-Julian value, the most energetic escaping particles can be
estimated using Eq.~(\ref{eqn22}). It will be $\sim \Delta \Phi /2$, or
\begin{eqnarray}
\epsilon_{\rm max} = 3 \times 10^{18}Z 
\left(  \frac{B_p}{10^{12} \rm G}  \right)
\left(  \frac{1 \rm ms}{P}  \right)^2
\left(  \frac{R}{10^6 \rm cm}  \right)^3   
\; \rm eV,
\label{eqn23}
\end{eqnarray} 
where $Z$ is the atomic charge. Thus the bulk Lorenz factor of protons
can be estimated as
\begin{eqnarray}
\Gamma_{\rm max} = 3.2 \times 10^9
\left(  \frac{B_p}{10^{12} \rm G}  \right)
\left(  \frac{1 \rm ms}{P}  \right)^2
\left(  \frac{R}{10^6 \rm cm}  \right)^3. 
\label{eqn24}
\end{eqnarray} 
In this study, we assume that the bulk Lorenz factor of the pulsar
wind is monolithic and its upper limit is given by Eq.~(\ref{eqn24}). 

\begin{figure}
\plotone{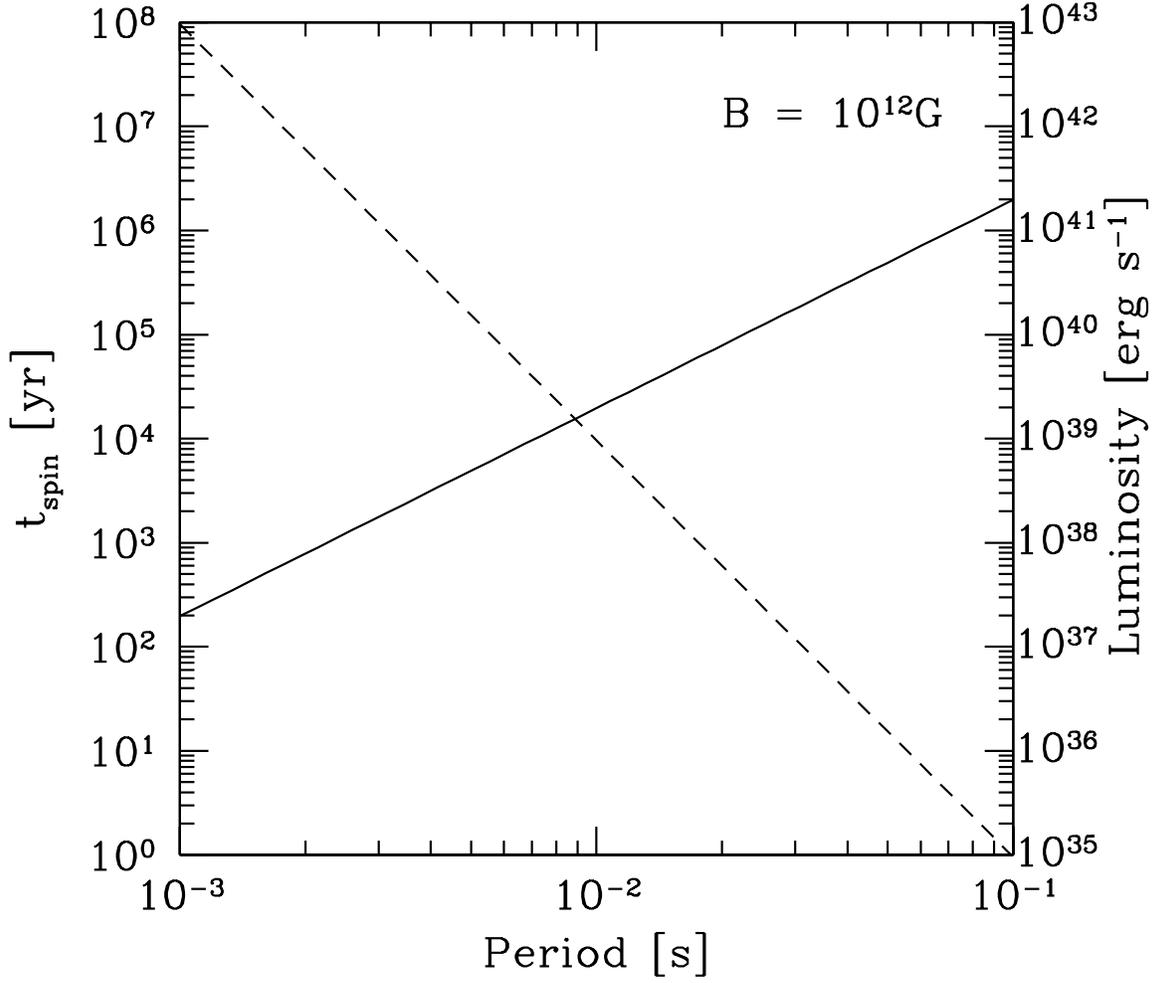}
\caption{
Solid line: spin down age [yr] as a function of period of a pulsar.
Spin down age is defined as the time interval for a pulsar's angular
velocity to become half of the initial one. Dashed line: spin down
luminosity as a function of a period of a pulsar. In this study,
wind luminosity is assumed to be comparable with the spin down luminosity.
Amplitude of the magnetic field at the pole of a pulsar is set to be
10$^{12}$G throughout in this study.
\label{fig1}}
\end{figure}

The wind luminosity can be estimated when it is assumed to be comparable
to the pulsar's spin down luminosity. Under the assumption,
it can be expressed as
\begin{eqnarray}
%\nonumber
L =
9.6 \times 10^{42} 
\left(   \frac{B_p}{10^{12} \rm G}   \right)^2
\left(   \frac{R}{10^{6} \rm cm}   \right)^6
\left(   \frac{1 \rm ms}{P}   \right)^4 \; \rm erg \; s^{-1}.
\label{eqn25}
\end{eqnarray}
In this case, the angular velocity evolves as a function of time as
\begin{eqnarray}
\Omega (t) = \Omega_i \left( 1 + \frac{B^2 R^6 \Omega_i^2}{3c^3 I} t
\right)^{-1/2} \;\; \rm rad \; s^{-1},
\label{eqn25-2}
\end{eqnarray}
where $\Omega_i$ and $I \sim 10^{45}$ g cm$^2$
are initial angular velocity and inertial
moment of a pulsar, respectively. We can estimate the spindown age
which is defined as the time when
the angular velocity $\Omega (t)$ becomes $2 \Omega_i$ as
\begin{eqnarray}
t_{\rm spin} \equiv 6.2 \times 10^9 \left(  \frac{10^{12} \rm G}{B_p}
  \right)^2
\left(  \frac{10^{6} \rm cm}{R}  \right)^6
\left(  \frac{P}{1 \rm ms}  \right)^2 \;\; \rm s.
\label{eqn25-3}
\end{eqnarray}
In figure~\ref{fig1}, we show the spin down age [yr] as a function of period
of a pulsar (solid line) as well as the spin down luminosity (dashed line).
Amplitude of the magnetic field at the pole of a pulsar is set to be
10$^{12}$G in this study.

\placefigure{fig1}

\begin{figure}
\plotone{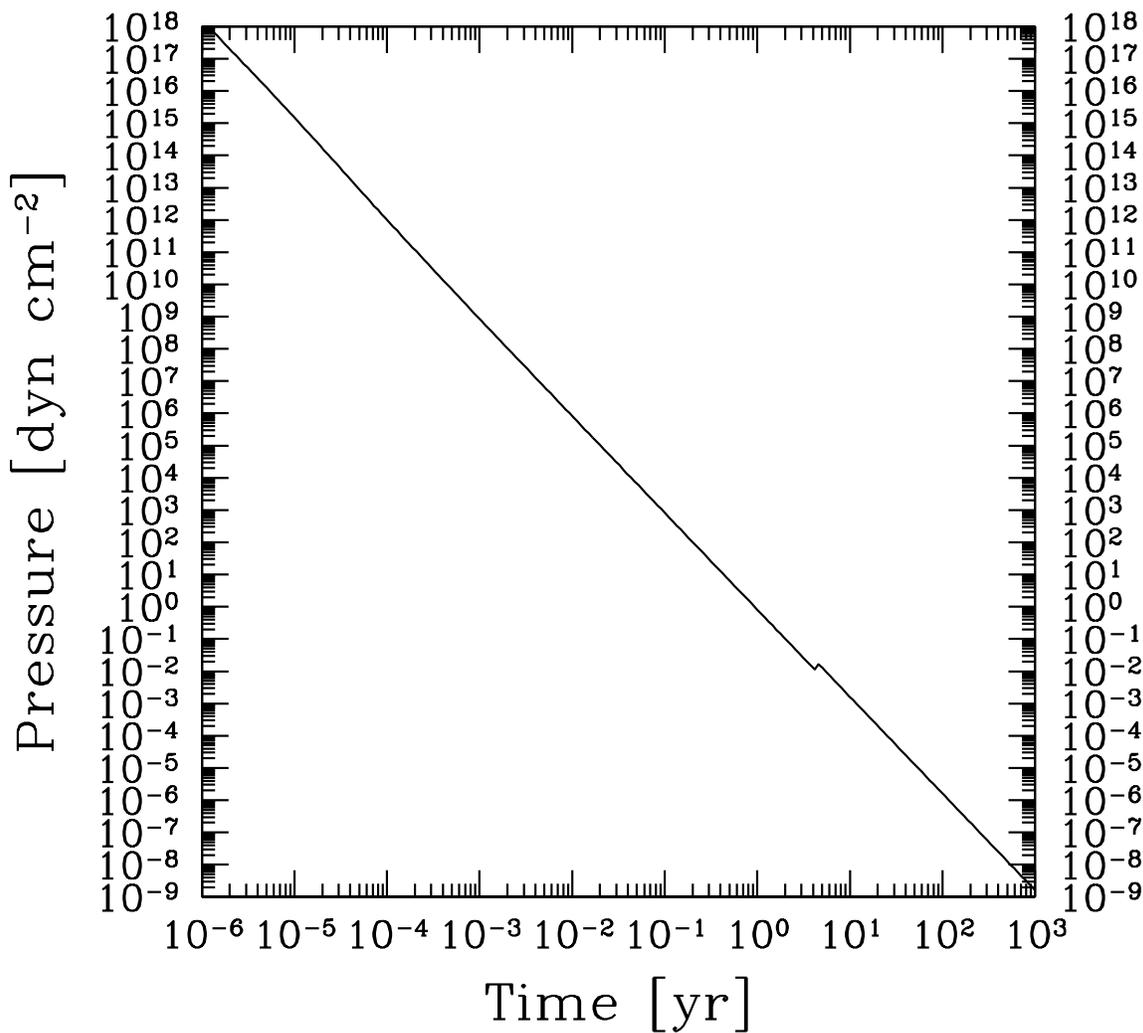}
\caption{
Relation between pressure [dyn cm$^{-2}$] and age [yr] of the
innermost region of the remnant. The discontinuity at $t \sim 5$ yr
reflects the transition from photon-dominated phase to matter-dominated phase.
This happens when the optical depth of the supernova ejecta becomes lower
than unity and pressure of photon fields is set to be zero.
\label{fig2}}
\end{figure}

Next, we consider the outer boundary condition.
It is naturally considered that the inner region of a supernova
remnant is composed of He layer that includes heavy
nuclei~\citep{hashimoto95,woosley95,thielemann96,nagataki97,nagataki00} whose
escape velocity is $\sim 2000$ km s$^{-1}$ for the free-expansion
phase~\citep{haas90,spyromilio90}. In this phase, the initial explosion
energy is almost all manifested by kinetic energy of the ejected matter;
thermal energy comes to 2 or 3$\%$ of the initial explosion energy.
Thus we model the innermost region of the ejecta as follows. We consider
the 6$M_{\odot}$ He layer which is the typical mass of helium
for the progenitor
of collapse-driven supernova~\cite{hashimoto95}. Then, the speed of the
escaping velocities of the He layer is 2000 km s$^{-1}$ (=$V_{\rm min}$)
for the inner edge
and 3000 km s$^{-1}$ (=$V_{\rm max}$)
for the outer edge. The thermal energy in the He layer is assumed to be 
\begin{eqnarray}
\nonumber
E_{\rm th} &=& 0.02 \times 10^{51} \times \frac{6M_{\odot}}{20M_{\odot}} \;
\rm erg \\
&=& 6 \times 10^{48} \; \rm erg,
\label{eqn26}
\end{eqnarray}
where 0.02 represents the fraction of the thermal energy in the ejecta,
$10^{51}$ is the typical explosion energy of a collapse-driven supernova,
$6M_{\odot}$ is the mass of the He layer and $20M_{\odot}$ is the typical
total mass of the ejecta. Thus the volume occupied by the He layer can be
calculated as
\begin{eqnarray}
V = \frac{4}{3}\pi \left[ V_{\rm max}^3  - V_{\rm min}^3\right] 
\left(  \frac{t}{1 \rm sec}  \right)^3,
\label{eqn27}
\end{eqnarray}
where $t$ is the age of the supernova remnant. Thus the typical pressure
in the He layer can be estimated by solving the equations
\begin{eqnarray}
E_{\rm th} = \frac{3}{2}(N_{\rm e} + N_{\rm He})k_B T + 3aT^4 V
\label{eqn28}
\end{eqnarray}
and
\begin{eqnarray}
P = (n_{\rm e} + n_{\rm He})  k_B T + aT^4,
\label{eqn29}
\end{eqnarray}
where $N_{\rm e}$ and $N_{\rm He}$ are the total number of electron and
helium in the He layer,
$n_{\rm e}$ and $n_{\rm He}$ are the number density of electron and helium, 
and $a$ is the radiation constant, respectively. Here we approximated that the
He layer is composed of helium, neglecting the contamination of heavy nuclei.
Using this model, we can estimate the pressure at the innermost region as
a function of time and the result is shown in figure~\ref{fig2}. The
discontinuity at $t \sim 5$ yr reflects the transition from photon-dominated
phase to matter-dominated phase.
This happens when the optical depth of the supernova ejecta becomes lower
than unity and pressure of photon fields is set to be zero.

\placefigure{fig2}

As for the Crab nebula (age of the Crab is $\sim 1000$yr),
the pressure at the innermost region of the remnant is estimated to
be in the range (1-10)$\times 10^9$ dyn cm$^{-2}$~\cite{kennel84a}. 
On the other hand, Iwamoto et al. (1997) estimated the pressure of 
the innermost region of the ejecta numerically and reported that
$P \sim 10^{16}$ dyn cm$^{-2}$ at $t = 100$ sec ($\sim 3 \times 10^{-6}$yr).
We can find from figure~\ref{fig2} that the estimation by Eq.~(\ref{eqn29})
reproduces these values fairly well. Thus we adopt this formula
in this paper throughout.

\begin{figure}
\plotone{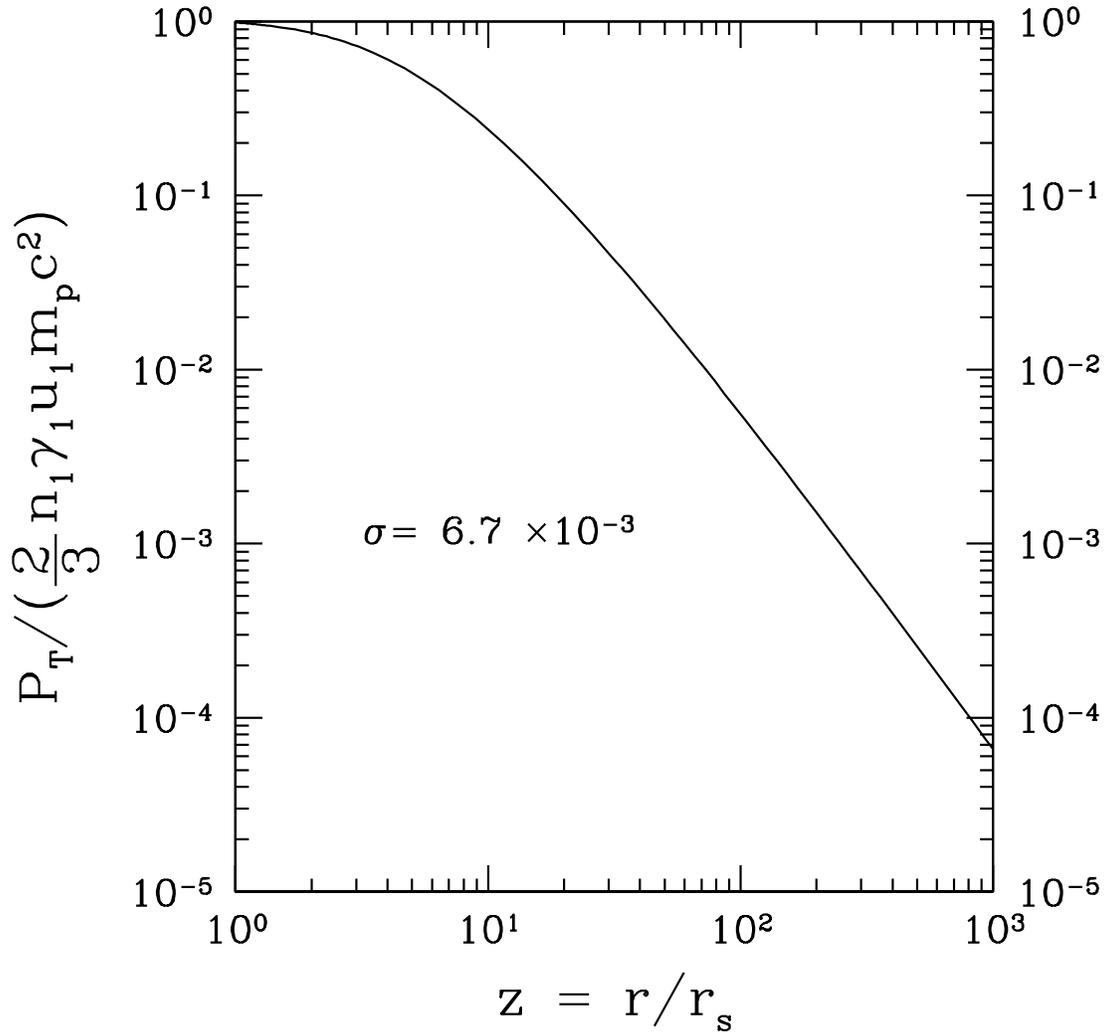}
\caption{
Total pressure as a function of $z=r/r_s$, normalized by
$2n_1\gamma_1u_1m_pc^2/3$. $\sigma$ is set to be
$6.7 \times 10^{-3} \equiv  \sigma_c$. 
\label{fig3}}
\end{figure}

Finally, we consider how to determine the position of the termination
shock. The termination shock should be initiated at the innermost region
of the remnant where the relativistic pulsar wind hits. Then, if the
stationary state exists, the position of the termination shock propagates
inward so as to attain the pressure balance between the nebula flow and
remnant. If the pressure of the nebular flow ($P_n$) can not be comparable
to that of the remnant ($P_R$), what happens?
If (a) $P_n > P_R$, the nebula flow should push the remnant until
the pressure balance is attained. In such a situation, interactions
between relativistic protons in nebula flow and non-relativistic protons
in the ambient remnant should be effective and may result in effective
production of high energy neutrinos. This situation will be similar
to that of the previous works such as Bednarek and Protheroe (1997).
In such a situation, the interactions
between the relativistic protons in the nebula flow and photons in the remnant
may also become important, which will be similar to the situation that Beall and
Bednarek (2002) presented. Such a situation is outscope of this study
and detailed estimation of flux of high-energy neutrinos has been done
by the previous works. On the other hand, if (b) $P_n < P_R$, the reverse shock
initiated at the inner edge of the remnant would be driven back to the
pulsar, as pointed out by Kennel and Coroniti (1984). This situation
is within our scope and considered. These situations can be distinguished by
using Eq.~(\ref{eqnadd9}). The total pressure normalized by
2$(n_1\Gamma_1u_1m_pc^2)/3 \sim P_2$ can be solved as a monotonic
function of $z$, as shown in figure~\ref{fig3}. In this figure, $\sigma$
is set to be $6.7 \times 10^{-3} \equiv  \sigma_c$,
which is used throughout this paper
as mentioned below. From the definition, $z_{\rm min} = 1$ and $z_{\rm max} =
(R_{\rm rem}/r_s)$, where $R_{\rm rem} = 2\times 10^8 t$ is the radius
of the inner edge of the remnant at time = $t$ sec.
If the pressure in the remnant ($P_R$) is
in the range $P_{\rm T}(z=z_{\rm max}) \le P_R \le P_{\rm T}(z=z_{\rm min})$,
the radius of the termination shock is determined at $r_s$ ($R \le r_s \le
R_{\rm rem}$). Otherwise, the situation is (a) $P_n > P_R$ or (b) $P_n < P_R$. 

\placefigure{fig3}

As for the velocity, the radial four velocity behind the shock
wave depends on only one parameter, $\sigma$ (Eq.~(\ref{eqn11})).
Also, in the small $\sigma$ limit, the asymptotic solution for
$u_{r \rightarrow \infty}$ can be written as
\begin{eqnarray}
u_{\infty} = \left(    \frac{\sigma^2}{1 + 2 \sigma}    \right)^{1/2},
\label{eqn30}
\end{eqnarray}
which corresponds to
\begin{eqnarray}
\beta_{\infty} = \frac{\sigma}{1 + \sigma}.
\label{eqn31}
\end{eqnarray}
Thus $\sigma$ should be $\sim 6.67 \times 10^{-3}$ in order to 
obtain the stationary nebula flow in which the velocity of the nebula
flow is almost same with that of the innermost region of the remnant. 
In this study, we set $\sigma = \sigma_c$ throughout in this study for
simplicity.

Of course, it is noted that the value of $\sigma$ should be not determined by
the outer boundary condition, but by pulsar's condition in principle. Thus
our requirement, which is also adopted in Kennel and Coroniti (1984),
may not be so realistic.
Thus, before we go further, we should discuss here what will happen
if (i) $\sigma > \sigma_c$ or (ii) $\sigma < \sigma_c$.
In the case of (i), the speed of the nebula flow is faster than that
of the remnant and additional pressure (ram pressure) is given to the
nebula flow. Thus discussion on the pressure balance mentioned above
should be modified by introducing the effect of the ram pressure
when the ram pressure is comparable with or higher than the total pressure. 
In the case of (ii), as long as the contact discontinuity is
adopted as a boundary condition, the location of the shock can not be
far away from the remnant so that the nebula flow can catch up with the
remnant. In this case, the resulting neutrino flux in the nebula flow
should be small because the volume of the cavity (i.e. region between
the shock and remnant) should be kept small. Also, there is a
possibility that shocked region and remnant continue to separate
from each other. If so, the rarefaction wave will be generated
from the remnant and the rarefaction wave will connect the remnant
and shocked region, achieving the boundary condition of contact
discontinuity between rarefaction wave and shocked region. It will be
important to investigate what kind of boundary condition is the most
proper as a next step of this study.

%---------------------------------------------------------------
\subsection{Microphysics of Proton-Proton Interaction}\label{micro}
%---------------------------------------------------------------
In this section, the formulation of production of neutrinos and gamma-rays
from $pp$ collisions is presented. 
In section~\ref{pion}, we calculate at first the pion spectrum in the rest
frame of the nebula flow. Then, the spectrum is transfered to the observer's
frame. In section~\ref{cross}, microphysics of $pp$ interactions is explained.
Finally, we calculate the flux of neutrinos and gamma-rays as
products of pion decays in section~\ref{neutrino}.

%-----------------------------------------------------------
\subsubsection{Formulation of Pion Production}\label{pion}
%-----------------------------------------------------------

First, we consider the number spectrum [particles cm$^{-3}$ s$^{-1}$
erg$^{-1}$] of charged and neutral pions in the rest frame of the nebula
flow. It is calculated by considering two protons with four momenta
$p_1$ and $p_2$, moving towards each other with relative velocity
$v_{\rm rel}$. The number of collisions that occur in a volume $dV$,
for a time $dt$, is a frame invariant quantity, which
in an arbitrary reference frame can be written as~\citep{landau75,mahadevan97}

\begin{eqnarray}
\nonumber
dR_{12} &=& \sigma_{pp} v_{\rm rel} \frac{cp_1}{E_1} \frac{cp_2}{E_2} n_1n_2dVdt
\\ 
&=& c \sigma_{pp} n_1n_2 \sqrt{(\vec{\beta_1} - \vec{\beta_2}  )^2-(\vec{\beta_1}
\times \vec{\beta_2})^2}dVdt, 
\label{eqn32}
\end{eqnarray}
where $\sigma_{pp}$ is the total cross section, $n_1$ and $n_2$
are the number density of protons with their four momenta are $p_1$
and $p_2$, respectively.
Thus the number spectrum of pions [particles cm$^{-3}$ s$^{-1}$ erg$^{-1}$]
is calculated as
\begin{eqnarray}
\nonumber
\frac{F(E_{\pi})}{dV} &=& c \int_1^{\infty} d \gamma_2
\int_1^{\gamma_2} d \gamma_1 \int^{1}_{-1} d \cos \theta  
\frac{d \sigma_{pp}(\gamma_1, \gamma_2, \cos \theta)}{dE_{\pi}}
n(R, \gamma_1) n(R, \gamma_2) \\
&\times& \sqrt{(\vec{\beta_1} - \vec{\beta_2} )^2 - (\vec{\beta_1} \times
\vec{\beta_2})^2},
\label{eqn33}
\end{eqnarray}
where $\gamma_1$, $\gamma_2$ are the respective Lorenz factors of the two
protons, $\cos \theta = \vec{\beta_1} \cdot \vec{\beta_2} / \left| 
\vec{\beta_1} \right|  \left| \vec{\beta_2} \right|$, $R$ is the radius
with respect to the neutron star, $n(R,\gamma)$ is the differential number
density of protons at position $R$, and $d \sigma_{pp}
(\gamma_1, \gamma_2, \cos 
\theta)/ d E_{\pi}$ is the differential cross section of proton-proton
interaction, which is explained in section~\ref{cross}. It is noted that
the more energetic proton is labeled as 2 and the less energetic one is
labeled as 1 in this formulation, which means $\gamma_2$ is always larger
than $\gamma_1$.
%The highest value for $\gamma_1$ is expressed as $\gamma$,
%which is the lowest value for $\gamma_2$ at the same time.
%Of course, the allowed range for $\gamma$ is [$1,\infty$].
In this frame, the distribution of pions in momentum space is isotropic.
Thus the number spectrum of pions in unit solid angle 
[particles s$^{-1}$ erg$^{-1}$ sr$^{-1}$] is simply expressed as
\begin{eqnarray}
\frac{dF(E_{\pi})}{d\Omega} = \frac{1}{4\pi} \int_{\Delta V}
\frac{F(E_{\pi})}{dV}  dV,
\label{eqn34}
\end{eqnarray}
where $\Delta_{V}$ is the fluid element.

Next, we transfer the obtained number spectrum in the fluid-rest frame
to the one in the observer's frame. It is noted that we consider a
receiver in the observer's frame. Thus we consider both special
relativistic effect and Doppler effect. The relative four velocity between
the nebula flow and observer is considered to be $u(r)$ (i.e. we
neglect the effect of proper motion of the pulsar with respect to the earth).
The result is expressed as (see Appendix~\ref{appb} for derivation)
\begin{eqnarray}
\frac{d\bar{F}(\bar{E}_{\pi}  )}{d\bar{\Omega}} = \frac{1}{\bar{\Gamma}^{2}
(1-\bar{\beta} \cos \bar{\theta})^2}\frac{F(E_{\pi})}{4\pi},
\label{eqn35}
\end{eqnarray}
where bars are labeled for the quantum in observer's frame,
$\bar{\theta}$ is the angle between the three dimensional velocity of
the fluid element and direction of the observer measured from the
fluid element,
and $\bar{\Gamma}$ is the bulk Lorenz factor of the fluid element in the
observer's
frame. The relation between $\bar{E}_{\pi}$ and $E_{\pi}$ is expressed as
\begin{eqnarray}
\bar{E}_{\pi} = \frac{1}{\bar{\Gamma} (1- \bar{\beta} \cos \bar{\theta}  )} E_{\pi}.
\label{eqn36}
\end{eqnarray}
This formulation is also used to calculate the flux of neutrinos and
gamma-rays in the observer's frame. In this study, however, this
transformation is not so effective except for the region just behind the
termination shock, because the speed of the flow behind the shock is
not so high in the case of $\sigma \ll 1$ (see Eqs.~(\ref{eqnadd7}) and
~(\ref{eqn30})).

%-------------------------------------------------------------------
\subsubsection{Differential Cross Section of Proton-Proton Interaction}\label{cross}
%-------------------------------------------------------------------

We show the method of calculating the differential cross section of
proton-proton interaction, $d \sigma_{pp} (\gamma_1, \gamma_2, \cos 
\theta)/ d E_{\pi}$, which is introduced in section~\ref{pion}.
We estimate this quantum by two steps. First, we transform the fluid-rest
frame to the rest frame of particle 1 (i.e. laboratory frame), because
formulation of the differential cross section for pion production in this
frame is presented by a number of previous
works~\citep{badhwar77,stephens81,dermer86a}. Using this formulation,
we can estimate the probability of producing pions with four momentum
$p^{'}_{\rm \pi}$ in this frame. Next, we transform again the coordinate
to the fluid-rest frame and obtain the differential cross section of
pion production in the fluid-rest frame. 

\begin{figure}
\plotone{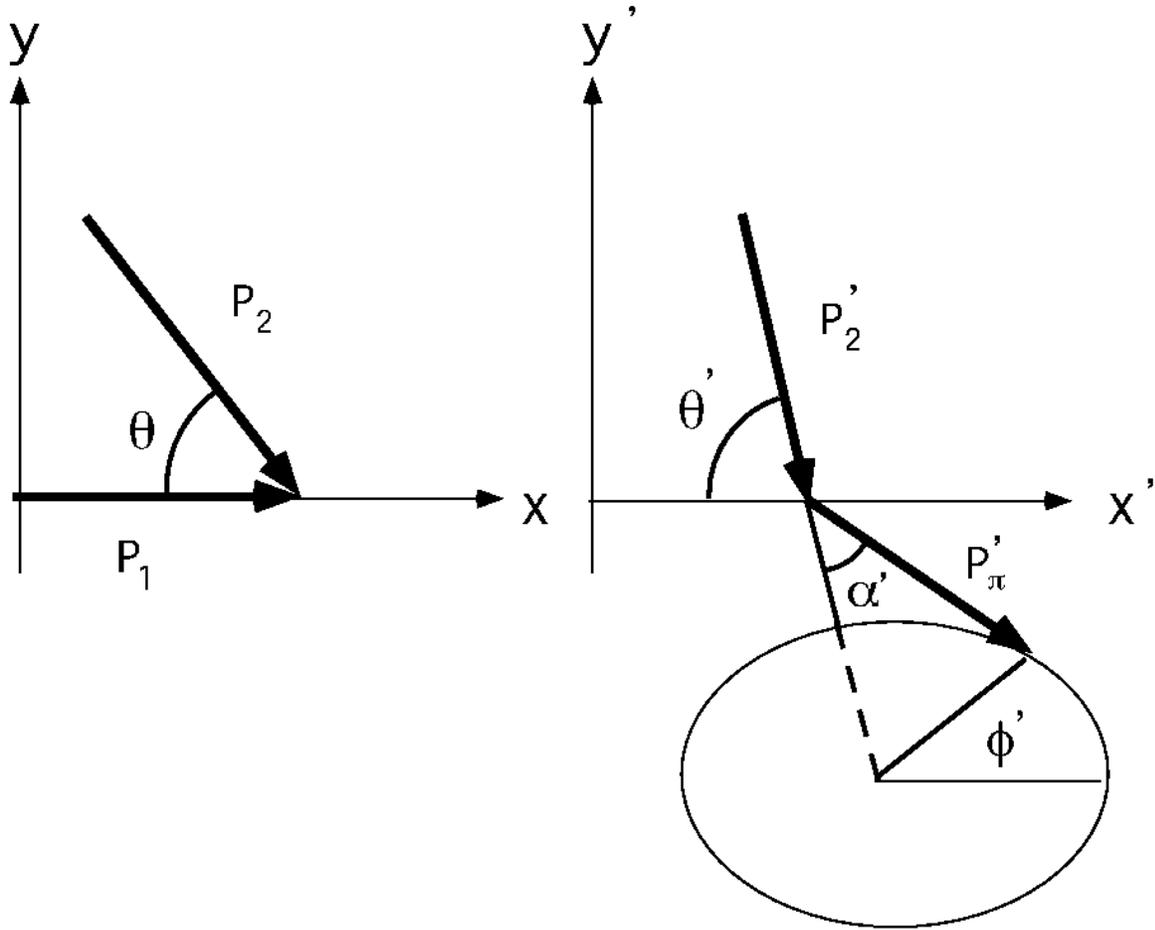}
\caption{
Sketch of the geometry concerning the individual scattering
events. The left panel shows the geometry in the fluid-rest frame,
while the right panel shows the geometry in the rest frame of particle 1.
\label{fig4}}
\end{figure}

Let us begin with transforming the fluid-rest frame to the rest frame
of particle 1. We choose $x-y$ plane so that the particles 1 and 2
moves in this plane. We also choose $x$ axis to be aligned with the
direction of velocity of particle 1. In fluid-rest frame, the particle 2
collide with particle 1 with angle $\theta$. In the rest frame of particle 1,
this angle becomes $\theta^{'}$. Pions are produced as a result of this
collision with zenith angle $\alpha^{'}$ and azimuthal angle $\phi^{'}$.
Here we choose $\alpha^{'} = 0$ axis to be aligned with the direction of
velocity of particle 2 in the rest frame of the particle 1. We also set
$\phi = 0$ and $\pi$ when the pion moves in the $x^{'}-y^{'}$ plane. 
We show a sketch of the geometry concerning the individual scattering
events in Figure~\ref{fig4}. The left panel shows the geometry in the
fluid-rest frame, while the right panel shows the geometry in the rest
frame of particle 1.

\placefigure{fig4}

In the fluid-rest frame, the four momentum of particle 2 can be expressed
as
\begin{eqnarray}
\left(
\begin{array}{cccc}
E_{2}/c \\
p_{2x}  \\
p_{2y}  \\
0
\end{array}
\right)
=
\left(
\begin{array}{cccc}
m_p c \gamma_2 \\
m_p c \sqrt{\gamma_2^2 -1} \cos \theta  \\
-  m_p c \sqrt{\gamma_2^2 -1} \sin \theta \\
0
\end{array}
\right).
\label{eqn37}
\end{eqnarray}
This is transformed to
\begin{eqnarray}
\nonumber
\left(
\begin{array}{cccc}
E_{2}^{'}/c \\
p_{2x}^{'}  \\
p_{2y}^{'}  \\
0
\end{array}
\right)
&=&
\left(
\begin{array}{cccc}
\gamma_1 & - \gamma_1 \beta_1 & 0 & 0 \\
- \gamma_1 \beta_1 &\gamma_1  & 0 & 0 \\
0 & 0 & 1 & 0 \\
0 & 0 & 0 & 1 \\
\end{array}
\right)
\left(
\begin{array}{cccc}
m_p c \gamma_2 \\
m_p c \sqrt{\gamma_2^2 -1} \cos \theta  \\
-  m_p c \sqrt{\gamma_2^2 -1} \sin \theta \\
0
\end{array}
\right)
\\ 
&=&
\left(
\begin{array}{cccc}
m_p c \gamma_1 (\gamma_2 - \beta_1 \sqrt{\gamma_2^2 -1}\cos \theta) \\
m_p c \gamma_1 (-\beta_1\gamma_2 + \sqrt{\gamma_2^2 -1}\cos \theta)  \\
-  m_p c \sqrt{\gamma_2^2 -1} \sin \theta \\
0
\end{array}
\right).
\label{eqn38}
\end{eqnarray}
Thus, the relation between $\theta$ and $\theta^{'}$ can be expressed as
\begin{eqnarray}
\tan \theta^{'} = - \frac{p_{2y}^{'}}{p_{2x}^{'}} = 
\frac{\sqrt{\gamma_2^2 -1}\sin \theta}{\gamma_1 \left( -\beta_1 \gamma_2 +
\sqrt{\gamma_2^2 -1} \cos \theta  \right)}
\label{eqn39}
\end{eqnarray}

In this frame, the differential cross section for pion production can
be expressed as\\
\cite{naito94}
\begin{eqnarray}
\frac{d \sigma_{pp}(E_{\pi}^{'},E_2^{'})}{dE_{\pi}^{'}} = 
\sqrt{E_{\pi}^{'2}- m_{\pi}^2 c^4}
\int_{0}^{2 \pi} d \phi^{'}
\int_{\cos \alpha_{\rm max}^{'}}^1 d \cos \alpha^{'}
\left(  E_{\pi}^{'}
\frac{d^3 \sigma_{pp}}{dp_{\pi}^{'3}c^3}  \right),
\label{eqn40}
\end{eqnarray}  
where $\cos \alpha_{\rm max}^{'} = \left[ \gamma_c E_{\pi}^{'} - E_{\rm max}^{*}(s) \right]/\left[  \beta_c \gamma_c p_{\pi}^{'} c  \right]$
($-1 \le \cos \alpha_{\rm max}^{'} \le 1$), $\gamma_c$
is the Lorenz factor of the center-of-mass system with respect to the rest
frame of particle 1, $E_{\rm max}^{*}(s) = 
(s - 4m_p^2 c^4 + m_{\pi}^2 c^4)/2s^{1/2}$ is the energy of pion in the
center-of-mass system when the pion obtains the maximum angle
$\alpha_{\rm max}^{'}$ in the rest frame of particle 1~\cite{dermer86b},
and $s$ is the Mandelstam variables for s-channel. 
The estimation of the Lorenz invariant quantum
$E_{\pi}^{'} d^3 \sigma_{pp}/dp_{\pi}^{'3}c^3$ is explained below.

The four momentum of produced pion in the rest frame of particle 1
can be expressed as 
\begin{eqnarray}
p_{\pi}^{'} =
\left(
\begin{array}{cccc}
\sqrt{p_{\pi}^{'2} + m_{\pi}^2 c^2} \\
p_{\pi}^{'} \{   \cos \alpha^{'} \cos \theta ^{'} - \sin \alpha^{'} \cos \phi^{'} \sin \theta^{'}   \}  \\
-p_{\pi}^{'} \{   \cos \alpha^{'} \sin \theta ^{'} + \sin \alpha^{'} \cos \phi^{'} \cos \theta^{'}   \}   \\
p_{\pi}^{'} \sin \alpha^{'} \sin \phi^{'}
\end{array}
\right).
\label{eqn41}
\end{eqnarray}
It can be transformed to the one in the fluid-rest frame as
\begin{eqnarray}
\left(
\begin{array}{cccc}
\gamma_1 &  \gamma_1 \beta_1 & 0 & 0 \\
 \gamma_1 \beta_1 &\gamma_1  & 0 & 0 \\
0 & 0 & 1 & 0 \\
0 & 0 & 0 & 1 \\
\end{array}
\right)
p_{\pi}^{'} =
\left(
\begin{array}{cccc}
\gamma_1 \sqrt{p_{\pi}^{'2} + m_{\pi}^2 c^2} + \gamma_1 \beta_1 p_{\pi}^{'}
\{   \cos \alpha^{'} \cos \theta ^{'} - \sin \alpha^{'} \cos \phi^{'}
\sin \theta^{'}   \}  \\
\gamma_1 \beta_1\sqrt{p_{\pi}^{'2} + m_{\pi}^2 c^2}+ \gamma_1 p_{\pi}^{'}
\{   \cos \alpha^{'} \cos \theta ^{'} - \sin \alpha^{'} \cos \phi^{'}
\sin \theta^{'}   \}   \\
-p_{\pi}^{'} \{   \cos \alpha^{'} \sin \theta ^{'} + \sin \alpha^{'}
\cos \phi^{'} \cos \theta^{'}   \} \\
p_{\pi}^{'} \sin \alpha^{'} \sin \phi^{'}
\end{array}
\right).
\label{eqn42}
\end{eqnarray}

Thus, since 
\begin{eqnarray}
\int d E_{\pi}  \frac{d \sigma_{pp} (\gamma_1, \gamma_2, \cos 
\theta)}{d E_{\pi}} =
\int d E_{\pi}^{'} 
\frac{d \sigma_{pp}(E_{\pi}^{'},E_2^{'})}{d E_{\pi}^{'}},
\label{eqn43}
\end{eqnarray}
$d \sigma_{pp} (\gamma_1, \gamma_2, \cos 
\theta)/d E_{\pi}$ can be expressed as
\begin{eqnarray}
\nonumber
\frac{d \sigma_{pp} (\gamma_1, \gamma_2, \cos \theta)}{d E_{\pi}}
= 
\frac{d}{dE_{\pi}}     \int \int \int 
dE_{\pi}^{'} d \phi^{'}  d \cos \alpha^{'} \sqrt{E_{\pi}^{'2}- m_{\pi}^2 c^4} 
\left(  E_{\pi}^{'} \frac{d^3 \sigma_{pp}}{dp_{\pi}^{'3}c^3}  \right) 
\\ \nonumber
\times H( \gamma_1 \sqrt{p_{\pi}^{'2} + m_{\pi}^2 c^2} 
+ \gamma_1 \beta_1 p_{\pi}^{'} \{   \cos \alpha^{'} \cos \theta ^{'} 
- \sin \alpha^{'} \cos \phi^{'} \sin \theta^{'}   \} - E_{\pi}) \\
\times H( E_{\pi} + dE_{\pi} - \gamma_1 \sqrt{p_{\pi}^{'2} + m_{\pi}^2 c^2}
+ \gamma_1 \beta_1 p_{\pi}^{'} \{   \cos \alpha^{'} \cos \theta ^{'} 
- \sin \alpha^{'} \cos \phi^{'} \sin \theta^{'}   \}) \},
\label{eqn44} 
\end{eqnarray}
where $H(x)$ is the Heaviside function.

Here we explain the formulation to estimate the Lorenz invariant cross
section $E_{\pi}^{'} d^3 \sigma_{pp}/dp_{\pi}^{'3}c^3$.
In the center-of-mass system (labeled as $*$),
the invariant cross section is inferred to have a form~\citep{badhwar77,naito94}
\begin{eqnarray}
E_{\pi}^{*} \frac{d^3 \sigma_{\pi}^{*}}{d^3 p_{\pi}^{*}} =
\frac{A}{(1+4m_p^2c^4/s)^r}(1-\tilde{x})^q \exp
\left[  \frac{Bp_{\bot}^{*}}{1+4m_p^2c^4/s}  \right],
\label{eqn45} 
\end{eqnarray}
where $\tilde{x} = \{ x_{\Vert}^{*2}  + (4/s)
(p_{\bot}^{*2}c^2 + m_{\pi}^2c^4)   \}^{1/2}$ and $q= (C_1 + C_2p_{\bot}^{*}
+ C_3p_{\bot}^{*} )$. Definition of $x_{\Vert}^{*}$ is $x_{\Vert}^{*} = 
p_{\Vert}^{*}c/ \sqrt{s/4 -m_p^2c^4}$. Here we decompose the momentum of pion
in the center-of-mass system of protons into the component which is parallel
to the velocity of the protons ($p_{\Vert}^{*}$) and the one which is
perpendicular to the velocity ($p_{\bot}^{*}$). The parameters
$A,B,C_1,C_2,C_3$ and $r$ for neutral and charged pions are tabulated in
table~\ref{tab1}. Note that the Lorenz invariant cross section is expressed in
units of mb/(GeV$^2$/c$^3$), momentum of pion is expressed in
units of GeV/c, and energy is expressed in units of GeV.
This fitting formula is derived from the study of $pp$ collisions
for incident proton energies 6-1500 GeV. Thus we have to note that
this formula has to be extrapolated to the high energy range in this
study. However, as shown in Naito and Takahara (1994), even if the scaling
model is extrapolate to higher energies, they confirmed that the scaling
assumption does not much affect the resulting gamma-ray spectrum by
comparing their results on the resulting gamma-ray spectrum up to
$\sim 10^7$ GeV with the gamma-ray production by the mini-jet model including
the QCD effect~\citep{gaisser87,berezinsky93}.
As for the low energy range ($\lesssim$ 5 GeV), the isobaric
model~\cite{stecker71}, in which a single pion is produced in a collision,
is considered to be better than the scaling model. However, we consider much
more energetic protons and the difference between the isobaric model and
the scaling model should be unimportant in this study.

\placetable{tab1}

%-----------------------------------------------------------
\subsubsection{Flux of Neutrino and Gamma-ray}\label{neutrino}
%-----------------------------------------------------------

The neutral and charged pions decay into gamma-rays, electrons (positrons),
and neutrinos as follows:
\begin{eqnarray}
\pi^{0} \rightarrow \gamma_1 + \gamma_2 \\
\pi^{+} \rightarrow \mu^{+} + \nu_{\mu} \rightarrow e^{+} + \nu_e 
+ \bar{\nu}_{\mu} + \nu_{\mu} \\
\pi^{-} \rightarrow \mu^{-} + \bar{\nu}_{\mu} \rightarrow e^{-} 
+ \bar{\nu}_e + \nu_{\mu} + \bar{\nu}_{\mu}.
\label{eqn46} 
\end{eqnarray}

In the case of 2-body decay ($\pi^{0} \rightarrow \gamma_1 + \gamma_2$
and $\pi^{\pm} \rightarrow \mu^{\pm} +  \nu_{\mu}(\bar{\nu}_{\mu})$),
the four momentum of gamma-ray and neutrino in the rest frame of the pion
can be obtained easily by calculating the conservation of four momentum.
As for the 3-body decay ($\mu^{\pm} \rightarrow e^{\pm} 
+ \nu_e(\bar{\nu}_e) + \bar{\nu}_{\mu}(\nu_{\mu})$), conservation of four
momentum and angular momentum (i.e. spin) have to be taken into account
to estimate the resulting energy spectrum of secondary neutrinos in the
rest frame of the charged pion. We adopt the formulation presented by
Dermer (1986a).
It is noted that the distribution of neutrino in the momentum space
is, after all, isotropic in the pion rest frame.
%It is noted that the direction of spin of a pion in its
%rest frame, the resulting momentum distribution of neutrino in the pion
%rest frame is isotropic in its frame.

Next, we transform the obtained
number spectrum of neutrino in the rest frame of pion to the one in the
fluid-rest frame. 
We can transform the number spectrum of neutrinos in pion-rest frame
to the one in fluid-rest frame by two steps, that is, Lorenz boost
and rotation. Let ($\theta$,$\phi$) be the zenith angle and azimuthal
angle of pion's velocity in the fluid-rest frame. We set $xyz$ coordinate
in the fluid-rest frame and $x^{''}y^{''}z^{''}$ in the pion-rest frame.
Here we choose $x^{''}$ axis to be parallel to the direction of pion's velocity.
When the four momentum of neutrino in the pion-rest frame is ($p_{\nu}^{''}$,
$p_{\nu}^{''} \cos \alpha^{''}$,  $p_{\nu}^{''} \sin \alpha^{''} \cos \beta^{''}$,
$p_{\nu}^{''} \sin \alpha^{''} \sin \beta^{''}$), it is transformed as
\begin{eqnarray}
\left(
\begin{array}{cccc}
E_{\nu}/c \\
p_{\nu x}  \\
p_{\nu y}  \\
p_{\nu z}
\end{array}
\right)
=
\left(
\begin{array}{cccc}
1 & 0 & 0 & 0 \\
0 & \sin \theta \cos \phi & \cos \theta \cos \phi & - \sin \phi \\
0 & \sin \theta \sin \phi & \cos \theta \sin \phi & \cos \phi   \\
0 & \cos \theta           & -\sin \theta          &  0          \\
\end{array}
\right)
\left(
\begin{array}{cccc}
\gamma & \gamma \beta & 0 & 0 \\
\gamma \beta & \gamma & 0 & 0 \\
0 & 0 & 1 & 0 \\
0 & 0 & 0 & 1 
\end{array}
\right)
\left(
\begin{array}{l}
p_{\nu}^{''} \\
p_{\nu}^{''} \cos \alpha^{''}  \\
p_{\nu}^{''} \sin \alpha^{''} \cos \beta^{''}  \\
p_{\nu}^{''} \sin \alpha^{''} \sin \beta^{''}
\end{array}
\right)
\label{eqn47}
\end{eqnarray}
Here $\alpha^{''}$ and $\beta^{''}$ are the zenith angle and
azimuthal angle of neutrino in the pion-rest frame. It is noted that
the zenith angle $\alpha^{''}$ is defined as the angle between the
velocity of neutrino and $x^{''}$ angle. $\gamma$ is the Lorenz factor
of pion in the fluid-rest frame. From Eq.~(\ref{eqn47}), the energy of
neutrino in the fluid-rest frame is expressed as
\begin{eqnarray}
E_{\nu}/c = \gamma p_{\nu}^{''}(1 + \beta \cos \alpha^{''}).
\label{eqn48}
\end{eqnarray}
Thus, by considering the conservation of number of neutrino in the fluid
element, the relation
\begin{eqnarray}
\frac{dN_{\nu}}{dE_{\nu}} dE_{\nu} = \int \int \int \int \int dp_{\nu}^{''} 
\frac{dn^{''}(p_{\nu}^{''})}{d p_{\nu}^{''}} 
\frac{d \beta^{''} \sin \alpha^{''} d \alpha^{''} }{4 \pi}
dp_{\pi} 4 \pi p^2_{\pi} \frac{d^3N^{\pi}}{dp^3_{\pi}}
dE_{\nu} \delta \{E_{\nu} - \gamma p_{\nu}^{''} (1+ \beta \cos \alpha^{''})c \}
\label{eqn49}
\end{eqnarray}
can be derived. Here $dN_{\nu}/dE_{\nu}$ is the number spectrum of
neutrino [particles GeV$^{-1}$] in the fluid-rest frame,
$dn^{''}(p_{\nu}^{''})/d p_{\nu}^{''}$ is the momentum distribution of a
neutrino in the pion-rest frame [(GeV/c)$^{-1}$], and $d^3N^{\pi}/dp^3_{\pi}$
is the number spectrum of pions in the momentum space in the fluid-rest
frame [particles (GeV/c)$^{-3}$].
Finally, the number spectrum of neutrino [particles GeV$^{-1}$] in the 
fluid element in the fluid-rest frame can be expressed as
\begin{eqnarray}
\frac{dN_{\nu}}{dE_{\nu}} = \frac{1}{2E_{\nu}} \int \int \int dp_{\nu}^{''} 
\frac{dn^{''}(p_{\nu}^{''})}{d p_{\nu}^{''}} 
\sin \alpha^{''} d \alpha^{''}
dp_{\pi} \frac{dN^{\pi}}{dp_{\pi}}
\delta \left\{
1 - \frac{\gamma p_{\nu}^{''} (1+ \beta \cos \alpha^{''})c}{E_{\nu}} 
\right\}.
\label{eqn50}
\end{eqnarray}
It is noted that the number spectrum of pions in the momentum space 
in the fluid-rest frame is isotropic so that it can be reduced to
the number spectrum of pion $dN^{\pi}/dp_{\pi}$ [particles (GeV/c)$^{-1}$].

Finally, we transform it to
the one in the observer's frame, using the formulation presented in
section~\ref{pion}.

%------------------------------------------------------------
\subsection{Detection of Neutrinos}\label{detection}
%------------------------------------------------------------

We briefly review the detectability of high energy neutrinos at km$^3$
detector such as IceCube.

In this study, we adopt the formulation presented by Gaisser and Grillo
(1987). The event rate [events s$^{-1}$] of the neutrino-induced signal
whose energy is larger than $E_{\mu}$
at the detector with effective area $A_{\rm eff}$ is estimated as
\begin{eqnarray}
S(\ge E_{\mu}) = A_{\rm eff} \int_{E_{\mu}}^{\infty} dE_{\nu} 
\frac{dN_{\nu}}{dE_{\nu}} P(E_{\nu},E_{\mu}),
\label{eqn51}
\end{eqnarray} 
where $dN_{\nu}/dE_{\nu}$ is the neutrino energy spectrum [particles cm$^{-2}$
s$^{-1}$ erg$^{-1}$] and $P(E_{\nu},E_{\mu})$ is the probability that a
neutrino aimed at the detector gives a muon with energy above $E_{\mu}$ at
the detector. The latter is given by
\begin{eqnarray}
P(E_{\nu},E_{\mu}) = \int_{E_{\mu}}^{E_{\nu}} dE_{\mu}^{\star} \int_{E_{\mu}^{\star}}^{E_{\nu}} dE^{\star \star}_{\mu} \frac{d \sigma}{dE^{\star \star}_{\mu}} N_A \int_0^{\infty}
dX g(X,E_{\mu}^{\star},E_{\mu}^{\star \star}),
\label{eqn52}
\end{eqnarray} 
where $N_A$ is the Avogadro's number and $g(X,E_{\mu}^{\star},E_{\mu}^{\star \star})$
is the probability that a muon produced with energy $E_{\mu}^{\star \star}$
travels a distance X g cm$^{-2}$ and ends up with energy in [$E_{\mu}^{\star},
E_{\mu}^{\star}+dE_{\mu}^{\star}$]. Gaisser and Grillo (1987) assumed that 
$g(X,E_{\mu}^{\star},E_{\mu}^{\star \star})$ has a form of
\begin{eqnarray}
g(X,E_{\mu}^{\star},E_{\mu}^{\star \star}) = 
\frac{\delta(X-X_0)}{\alpha (1+E_{\mu}^{\star \star}/\epsilon)}, \;\;\;
X_0 = \frac{1}{\beta} \ln \left[   
\frac{E_{\mu}^{\star \star}+\epsilon}{E_{\mu}^{\star}+\epsilon}    
\right],
\label{eqn53}
\end{eqnarray} 
where $\alpha$ = 2 MeV cm$^2$ g$^{-1}$, $\epsilon$ = 510 GeV,
and $\beta$ = 3.92$\times 10^{-6}$ cm$^2$ g$^{-1}$.
The cross section $d \sigma/dE^{\star \star}_{\mu}$ cm$^2$ erg$^{-1}$ can be
expressed by the differential cross section, which is explained below, as
\begin{eqnarray}
\frac{d \sigma}{dE^{\star \star}_{\mu}} = \frac{1}{E_{\nu}} \int_0^1
\frac{\partial ^2 \sigma}{\partial x \partial y}dx.
\label{eqn54}
\end{eqnarray}

The inclusive cross section for the reaction $\mu_{\nu} + N
\rightarrow \mu^{-}$ + anything is obtained by the
renormalization-group-improved parton model
($N=(n+p)/2$ is the isoscalar nucleon).
In this model, the differential cross section is
written in terms of the scaling variables $x= Q^2/2M \nu$ and
$y= \nu /E_{\nu}$ as
\begin{eqnarray}
\frac{\partial ^2 \sigma}{\partial x \partial y} = \frac{2G_F^2 M E_{\nu}}{\pi}
\left[ \frac{M_W^2}{Q^2 + M_W^2} \right]^2 \left\{  
xq(x,Q^2) + x(1-y)^2 \bar{q}(x,Q^2)
\right\},
\label{eqn55}
\end{eqnarray} 
where $-Q^2$ is the invariant momentum transfer between the incident neutrino
and outgoing muon, $\nu = E_{\nu} - E_{\mu}$ is the energy loss in the
laboratory frame, $M$ and $M_W$ are the nucleon and intermediate-boson masses,
and $G_F$ is the Fermi constant. As for the distribution function of valence
and sea quarks, we adopt the fitting formula (Set 2 with $\Lambda=290$ MeV)
presented by Eichten-Hinchliffe-Lange-Quigg (EHLQ) for
$x \ge 10^{-4}$~\citep{eichten84, eichten86}, while the double-logrithmic
approximation (DLA) presented by Reno and Quigg (1988) for
$x \le 10^{-4}$.

%%%%%%%%%%%%%%%%%%%%%%%%%%%%%%%%%%%%%%%%%%%%%%%%%%%%%%%%%%
\section{RESULTS}\label{results}
%%%%%%%%%%%%%%%%%%%%%%%%%%%%%%%%%%%%%%%%%%%%%%%%%%%%%%%%%%

In this study, when we estimate the flux of neutrinos and gamma-rays
from pion decays, we have to check whether the energy spectrum
of protons can be regarded to obey the relativistic Maxwellian distribution.
In the cases in which energy loss processes are so effective that 
the energy spectrum of protons do not obey the relativistic Maxwellian distribution
any longer, the formulation in this study can not be used and Boltzmann
equation should be needed to estimate the proton's energy distribution
and resulting flux of neutrinos, which is outscope and next step of this
study. Thus, we restrict the situations that meet the following constraints
in the {\it whole} region of the nebula flow. These are:
(i) production rate of pions [erg s$^{-1}$] is much smaller than
the luminosity of the pulsar wind. (ii) synchrotron cooling timescale of
protons is longer than traveling timescale (defined below) or timescale
of $pp$ collisions. (iii) energy transfer timescale from protons to electrons
is longer than traveling timescale or timescale of $pp$ collisions (see also
the footnotes of table 2). 

\placetable{tab2}

Here we introduce the definitions of the timescales mentioned above.
As for the synchrotron cooling timescale of protons can be written as
\begin{eqnarray}
t_{p, \rm syn} &=& \left( \frac{m_{p}}{m_e} \right)^4 \times
t_{e, \rm syn} \sim 1.1 \times 10^{13} t_{e, \rm syn}, 
\label{result-1}
\end{eqnarray}
where $t_{e, \rm syn}$ is the synchrotron energy loss timescale for
an electron and can be written as
\begin{eqnarray}
t_{e, \rm syn} = 3.9 \times 10 \left( \frac{1 \; {\rm GeV}}{E} \right)
\left(  \frac{10^2 \; {\rm G}}{B}  \right)^2 \;\;\; \rm s.
\label{result-2}
\end{eqnarray}
Traveling timescale, which means the timescale for the bulk flow of
the nebula flow to be conveyed to outer region, is defined as
\begin{eqnarray}
t_{\rm travel} = \frac{r}{v},
\label{result-3}
\end{eqnarray}
where $r$ is the location of the fluid element measured from the pulsar
and $v$ is the bulk speed of the nebula flow.
Collision timescale between protons is estimated as
\begin{eqnarray}
t_{\rm col} = \frac{1}{n \sigma_{pp}c},
\label{result-4}
\end{eqnarray}
where $n$ is the number density of protons, $\sigma_{pp}$ is the
cross section of $pp$ interaction, and $c$ is the speed of light.
For simplicity, to estimate this timescale, $\sigma_{pp}$ is set
to be 100mb.
Collision timescale of energy transfer from protons to electrons
is estimated as~\cite{stepney83}
\begin{eqnarray}
t_{\rm ep} = \frac{4}{\ln \Lambda} \frac{n_e}{n_p}\left( \frac{kT_e}{m_e c^2}
\right)^2 \frac{1}{n_e \sigma_T c}
\label{result-5}
\end{eqnarray}
where $\sigma_{T}$ is the Thomson cross section and $\ln \Lambda$ is the
value of the Coulomb logarithm which is estimated as
\begin{eqnarray}
\ln \Lambda = \ln \left[  \frac{k T_e}{\hbar \omega_p} \right].
\label{result-6}
\end{eqnarray}
Here $\omega_p$ is the plasma frequency and is given as
\begin{eqnarray}
\omega_p = \left( \frac{4 \pi e^2 c^2 n_e}{3kT_e}  \right)^{1/2}
\label{result-7}
\end{eqnarray}
for a highly relativistic thermal electron gas~\cite{gould81}.
Here we assume that $T_e = (m_e/m_p)T_p$ and $n_e=n_p$ to estimate the
energy transfer timescale, although we found that the energy transfer
timescale depends very weakly on their number ratio. In fact, we found that
the energy transfer timescale hardly changes even if the number density
of electrons becomes $\sim 10^3$ times larger than that of protons.

Some explanations should be added to the constraints (ii) and (iii).
We consider that the Maxwellian distribution is hold even if the
synchrotron cooling timescale of protons is shorter than the pion
production timescale, as long as the traveling timescale is shorter
than the synchrotron cooling timescale. This means the flow is conveyed
to outer region without losing their energy. It is noted that the definition
of the traveling timescale is also frequently used as the timescale of
adiabatic cooling. However, effect of adiabatic cooling has been already
taken into consideration because we adopt the formulation of nebula flow.
%However, this is the case where the amplitude of the
%coherent magnetic field is strong and charged particles is trapped in the
%field. On the other hand, in this study, the energy density of the magnetic
%field is much smaller ($\sigma \ll 1$ in section~\ref{form}) than the energy
%density of the chaged particles. Thus the magnetic field should be randomaized
%and chaged particles should not be trapped by the magnetic field at all.
%Thus we neglected in this study the effect of the adiabatic cooling
%and regard the $t_{\rm travel}$ as the traveling timescale.
Next, we found that the synchrotron cooling timescale of electrons is
shorter than the energy transfer timescale from protons to electrons
whenever the constraint (iii) is not hold. Thus, in the cases where
the constraint (iii) is not hold, most of the energy of protons is
considered to be lost through the synchrotron cooling of electrons. 
Finally, we assume that the energy distribution of protons is hold
to be Maxwellian. However, exactly speaking, we should consider
another condition, that is, whether the Coulomb collision timescale
between protons is sufficiently short that the Maxwellian distribution
can be hold. If this condition can not be satisfied, temperature of
protons can not be defined exactly and formulation of the nebula flow
must be altered by using Boltzmann equation. However, as explained
in section~\ref{nebula}, we found in this study that charged and neutral
pions are mainly produced just behind the termination shock where
the energy distribution of protons obey relativistic Maxwellian. 
Thus, we think that the order-estimate of the neutrino flux produced
by $pp$ collisions will be valid even if the MHD equations are adopted.

\begin{figure}
\plotone{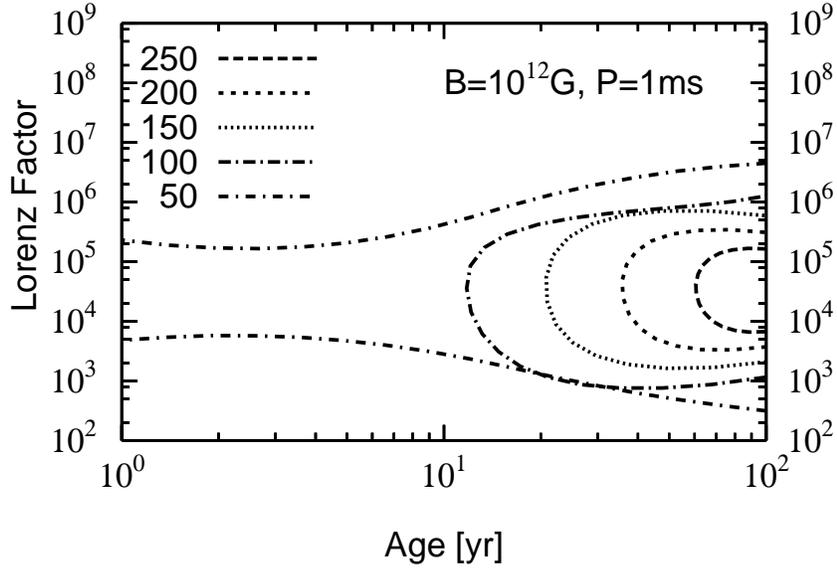}
\caption{
Contour of the neutrino event whose energy is greater than 10 GeV per year
for a km$^3$ detector of high-energy neutrinos as functions of age [yr]
of a pulsar and initial bulk Lorenz factor of the pulsar wind.
The pulsar is assumed to be located 10 kpc away from the earth.
The amplitude of the magnetic field and period of the pulsar are assumed to
be $10^{12}$ G and 1ms. It is noted that the spin down age is about
2$\times 10^2$ yr, so that the range of the age is limited to be within
$10^2$ yr. Also, when the age of the pulsar is about $10^{-1}$ yr,
synchrotron cooling timescale and/or energy transfer timescale from protons
to electrons are so short that the Maxwellian distribution of
protons is distorted and production of pions is severely limited (see table 2).
As for the bulk Lorenz factor of the initial pulsar wind, the maximum value
is determined by Eq.(28). When the bulk Lorenz factor is about $10^1$, the
energy transfer timescale from protons
to electrons are too short (see table 2).
\label{fig5}}
\end{figure}

Here we show the results.
At first, in figure~\ref{fig5}, we show contour of the neutrino event
whose energy is greater than
10 GeV per year for a km$^3$ detector of high-energy neutrinos as functions
of age [yr] of a pulsar and initial bulk Lorenz factor of the pulsar wind
(Models A2-A9, B2-B9, and C2-C9).
The pulsar is assumed to be located 10 kpc away from the earth. The amplitude
of the magnetic field and period of the pulsar are assumed to be $10^{12}$G
and 1ms. It is noted that the spin down age is about 2$\times 10^2$ yr, so
that the range of the age is limited to be within $10^2$ yr. Also, when the
age of the pulsar is about $10^{-1}$ yr (Models D4-D9), synchrotron cooling timescale
is so short that the condition (ii) is not satisfied (see table 2).
As for Model D3, collision timescale between protons is the fastest one and
the condition (i) is not satisfied.
As for the bulk Lorenz factor of the
initial pulsar wind, the maximum value is determined by Eq.(\ref{eqn24}). When the
bulk Lorenz factor is about $10^1$, the energy transfer timescale from protons
to electrons are too short (see table 2).

\placefigure{fig5}

\begin{figure}
\plotone{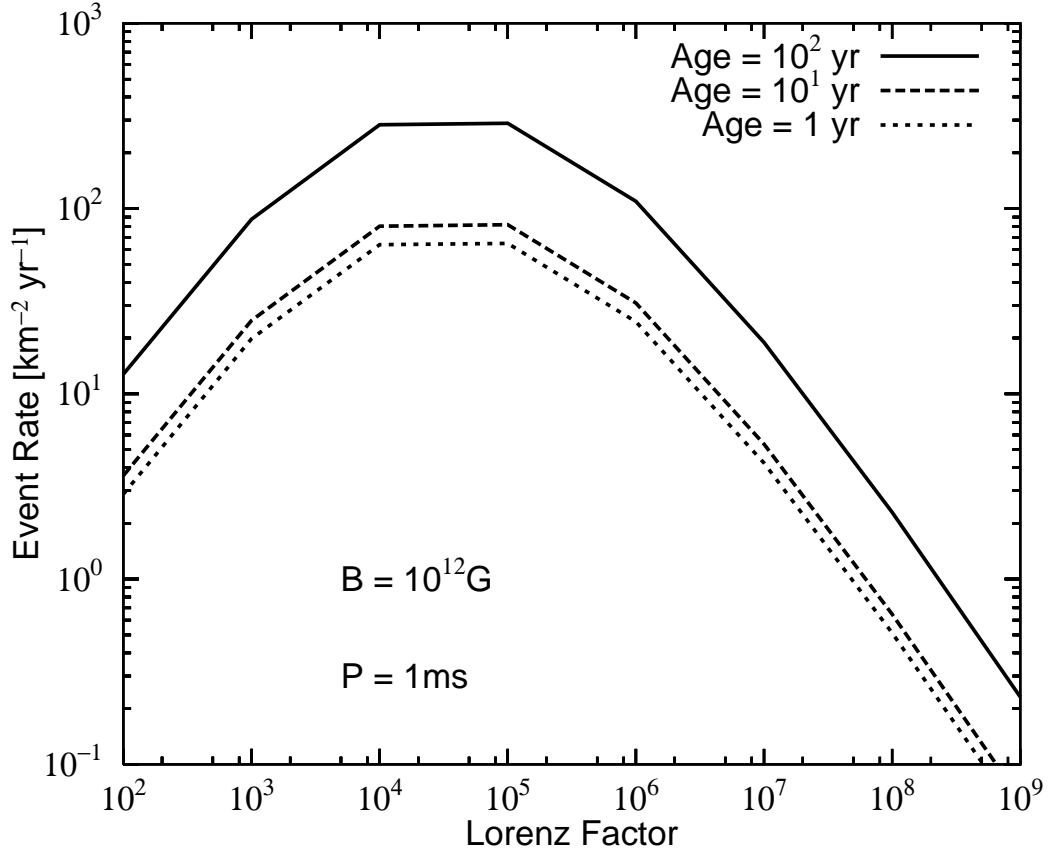}
\caption{
Same with figure 5, but horizontal line represents the initial bulk Lorenz
factor of pulsar wind with the age of the pulsar fixed as 1 (Models C2-C9),
$10^1$ (Models B2-B9), and $10^2$yr (Models A2-A9). The reason for
the event rate to become smaller for the case of large initial Lorenz factor
is that the number density of protons behind the termination shock wave
becomes smaller (see figure 7). Also, the reason for the event rate to become
smaller for the case of the Lorenz factor to be smaller than $10^4$ is that
high energy protons do not exist and high energy neutrinos are not produced.
It is noted that the cross section between neutrino and quark is roughly
proportional to energy of neutrinos in the quark-rest frame. 
\label{fig6}}
\end{figure}

\begin{figure}
\plotone{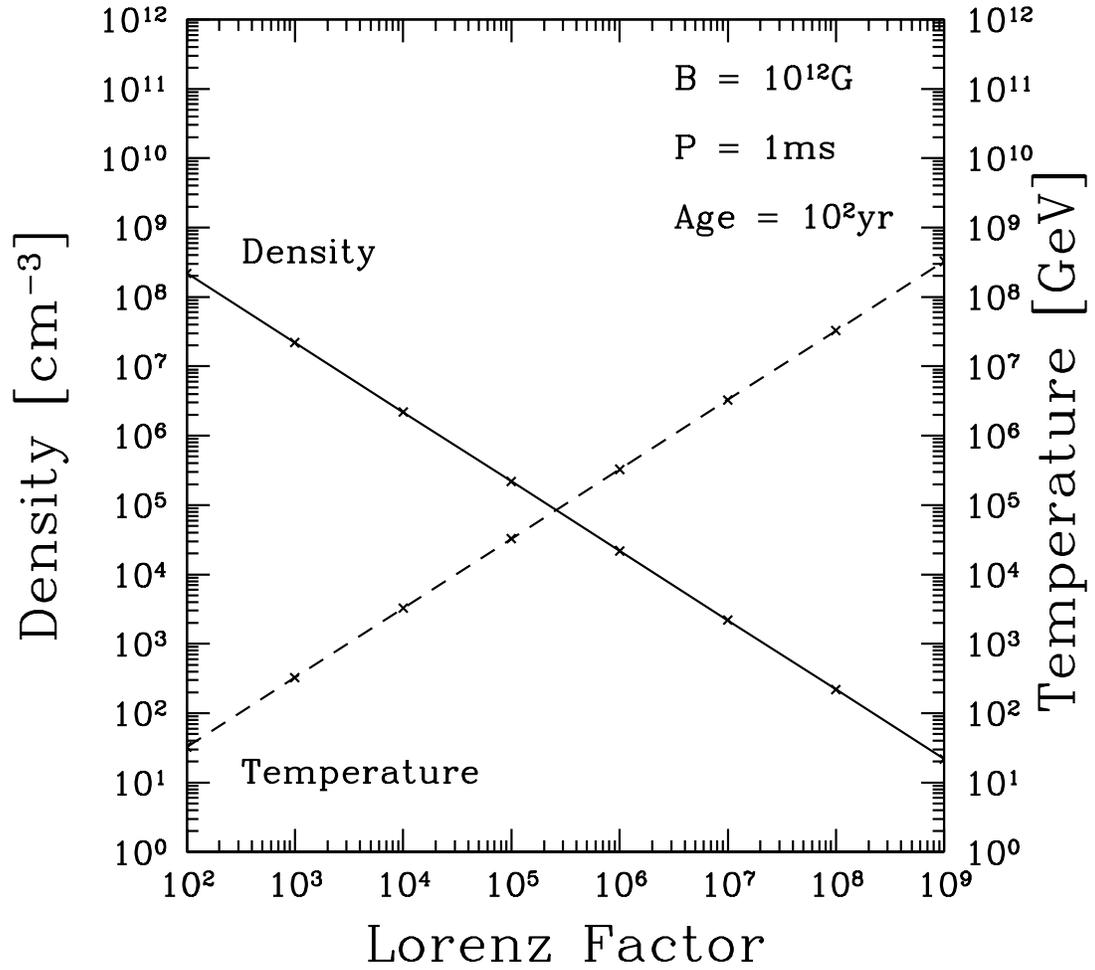}
\caption{
The density (solid line) and temperature (dashed line) behind the
termination shock wave as
a function of the initial bulk Lorenz factor of the pulsar wind.
The amplitude of the magnetic
field, period, and age of the pulsar are assumed to be $10^{12}$G, 1ms,
and $10^2$yr (Models A2-A9).  
\label{fig7}}
\end{figure}

Figure~\ref{fig6} is same with figure~\ref{fig5}, but horizontal line
represents the initial bulk Lorenz factor of pulsar wind with the age of the
pulsar fixed as 1 (Models C2-C9), $10^1$ (Models B2-B9), and $10^2$yr (Models A2-A9).
The reason for the event rate to
become smaller for the case of large initial Lorenz factor is, as shown
in figure~\ref{fig7}, that the number density of protons behind the termination
shock becomes smaller. Also, the reason for the event rate to become
smaller for the case of the Lorenz factor to be smaller than $10^4$ is that high energy
protons do not exist and high energy neutrinos are not produced. It is noted
that the cross section between neutrino and quark is roughly proportional to
energy of neutrinos in the quark-rest frame. 

\placefigure{fig6}

Figure~\ref{fig7} represents the density (solid line) and temperature
(dashed line) behind the termination shock as a function of the initial
bulk Lorenz factor of the pulsar wind. The amplitude of the magnetic
field, period, and age of the pulsar are assumed to be $10^{12}$G, 1ms,
and $10^2$yr (Models A2-A9). As mentioned above, the number density of protons is smaller
for the larger initial bulk Lorenz factor. We can also confirm that the
temperature of protons behind the termination shock wave is roughly same with
the initial kinetic energy of the protons. 

\placefigure{fig7}

\begin{figure}
\plottwo{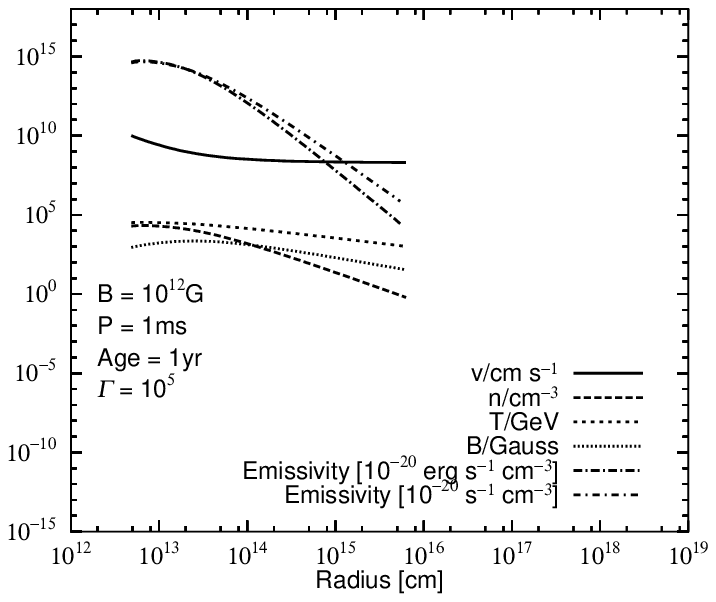}{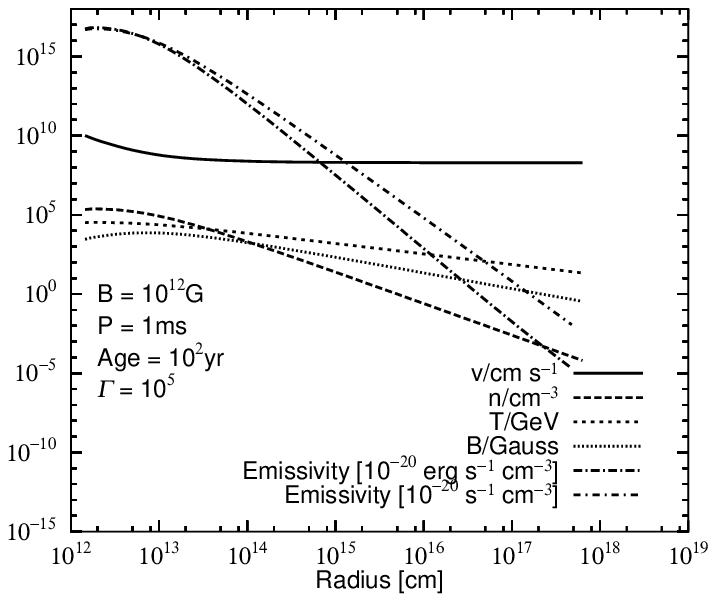}
\caption{
Profiles of velocity, number density of protons, temperature, magnetic
field, and emissivity of charged pions in units of
[10$^{-20}$erg s$^{-1}$ cm$^{-3}$] and
[10$^{-20}$particles s$^{-1}$ cm$^{-3}$]. The inner and outer boundaries
correspond to the location of the termination shock wave and the innermost
region of the surrounding supernova remnant.
The amplitude of the magnetic field and period of the pulsar are assumed
to be $10^{12}$G and 1ms. Initial bulk Lorenz factor of the pulsar wind
is set to be $10^5$.
The left panel represents the case that the age of the pulsar is 1 yr
(Models C5),
while the right panel shows the case that the age of the pulsar is
$10^2$ yr (Models A5).
\label{fig8}}
\end{figure}

The reason why the flux of neutrinos becomes larger with time can be
understood by figure~\ref{fig8}. Figure~\ref{fig8} shows profiles of
velocity, number density of protons, temperature, magnetic field,
and emissivity of charged pions in units of [10$^{-20}$erg s$^{-1}$
cm$^{-3}$] and [10$^{-20}$particles s$^{-1}$ cm$^{-3}$]. The inner and
outer boundaries correspond to the location of the termination shock wave
and the innermost region of the surrounding supernova remnant.
The amplitude of the magnetic field and period of the pulsar are assumed to
be $10^{12}$G and 1ms. The left panel represents the case that the age of the
pulsar is 1 yr (Model C5), while the right panel shows the case that the age of
the pulsar is $10^2$ yr (Model A5).
In both cases, initial bulk Lorenz factor of the pulsar wind is set to be
$10^5$.
The pressure of the innermost region of the remnant
becomes smaller with time (see figure~\ref{fig2}). Thus the location of the
termination shock moves inward with time, which results in higher density
and higher emissivity of neutrinos. Also, additionally, the radius of the
remnant becomes outward, which results in the region of the nebula flow
becomes larger with time. These are the reason why the flux of neutrinos
becomes larger with time. We can also find that the emissivity of charged
pions [particles cm$^{-3}$ s$^{-1}$] in the nebula flow can be roughly estimated as
\begin{eqnarray}
\epsilon = n^2 c \sigma_{pp} = 3 \times 10^{-5} \left(  \frac{n}{10^5
{\rm cm^{-3}}}
%\right)
\right)^2
\left(  \frac{\sigma_{pp}}{100 \rm mb} \right).
\label{result-8}
\end{eqnarray}
For example, in the left panel of figure~\ref{fig8}, the density and emissivity
just behind the termination shock are about 10$^5$ [cm$^{-3}$] and
$10^{-5}$ [particles cm$^{-3}$ s$^{-1}$], respectively.
Also, those values just forward the innermost region of the surrounding
supernova remnant are about 1 [cm$^{-3}$] and
$10^{-14}$ [particles cm$^{-3}$ s$^{-1}$], respectively. 
Thus the emissivity can be roughly reproduced by using Eq.~(\ref{result-8}),
although the emissivity is obtained by using formulation presented
in section~\ref{pion}.
This means the rough estimate holds a good approximation even if the protons
obey the Maxwellian distribution.
Also, as pointed out in section~\ref{nebula},
we can find that pions are mainly produced just behind the termination shock.

\placefigure{fig8}

Before we go to the further discussion, we consider the proper value for the
bulk Lorenz factor of the pulsar wind by introducing the Goldreich-Julian value,
which is usually used to estimate the charge density of the $e^+e^-$ pair plasma,
although the initial bulk Lorenz factor of the pulsar wind is treated
as a free parameter in figure~\ref{fig7}. Goldreich-Julian value at the
equatorial plane at radius $r$ can be written as
\begin{eqnarray}
n_{\rm GJ}(r) = \frac{\vec{B}(r) \cdot \vec{\Omega}}{2 \pi e c} = \frac{B \Omega}{2 \pi e c}
              = 6.9 \times 10^{-2} \frac{B(r)}{P} \;\;\; \rm cm^{-3},
\label{result-9}
\end{eqnarray} 
where $B(r)$ is the amplitude of the magnetic field at radius $r$ in a unit of G, $\Omega$ is the
angular velocity of a pulsar, and $P$ is the period of the pulsar in a unit of sec.
Assuming that the magnetic field around the pulsar is dipole, Eq.(\ref{result-9})
can be written as
\begin{eqnarray}
n_{\rm GJ}(r) = 6.9 \times 10^{13} \left( \frac{1 \rm ms}{P} \right) \left( \frac{B_p}{10^{12} \rm G} \right)
\left( \frac{10^6 \rm cm}{r} \right)^3  \;\;\; \rm cm^{-3}, 
\label{result-10}
\end{eqnarray} 
where $B_p$ is the amplitude of the magnetic field at the pole of the pulsar.
Since the location of the light cylinder is
\begin{eqnarray}
R_{\rm lc} = 4.8 \times 10^6 \left(   \frac{P}{1 \rm ms}     \right) \;\;\; \rm cm,
\label{result-11}
\end{eqnarray} 
the Goldreich-Julian value at the light cylinder can be expressed as
\begin{eqnarray}
n_{\rm GJ}(R_{\rm lc}) = 6.4 \times 10^{11} \left( \frac{1 \rm ms}{P} \right)^4 \left( \frac{B_p}{10^{12} \rm G} \right) \;\;\; \rm cm^{-3}.
\label{result-12}
\end{eqnarray} 
From this value, we can estimate the net number density of electrons ($\mid n(e^+) - n(e^-) \mid$)
in the pulsar wind just ahead of the termination shock
as
\begin{eqnarray}
n(r_s) =   \left(  \frac{R_{\rm lc}}{r_s} \right)^2 n_{\rm GJ} (R_{\rm lc}) \;\;\; \rm cm^{-3},
\label{result-13}
\end{eqnarray}
where we use the conservation law of number flux of protons, $4 \pi R_{\rm lc}^2 \Gamma n_{\rm GJ}(R_{\rm lc})
= 4 \pi r_{s}^2 \Gamma n(r_{s})$.
Eq.(\ref{result-13}) can be written by inserting Eq.(\ref{result-12}) as
\begin{eqnarray}
n(r_s) =   1.5 \times 10 \left( \frac{1 \rm ms}{P} \right)^2 \left( \frac{B_p}{10^{12} \rm G} \right)
\left(  \frac{10^{12}{\rm cm}}{r_s} \right)^2 \;\;\; \rm cm^{-3}.
\label{result-14}
\end{eqnarray}
The location of the termination shock is found to be $\sim 10^{12}$ cm from figure~\ref{fig8}.
On the other hand, the number density of protons can be estimated by using relation
$L \sim 4 \pi n \Gamma^2 r_s^2 m_p c^3$ (see Eqs.(\ref{eqn2}) and (\ref{eqn25}). note that $\sigma$ is much smaller than
unity and bulk flow is highly relativistic). It can be expressed as
\begin{eqnarray}
n(r_s) =   1.7 \times  \left( \frac{10^5}{\Gamma} \right)^2 \left(  \frac{10^{12}{\rm cm}}{r_s} \right)^2
\left( \frac{B_p}{10^{12} \rm G} \right)^2 \left( \frac{1 \rm ms}{P} \right)^4 \left( \frac{R}{10^6 \rm cm} \right)^6 \;\;\; \rm cm^{-3}. 
\label{result-15}
\end{eqnarray}
By comparing Eq.(\ref{result-14}) with Eq.(\ref{result-15}) and assuming that number density
of proton is comparable with the net number density of electrons ($\mid n(e^+) - n(e^-)   \mid $), proper value for
the bulk Lorenz factor seems to be $\Gamma \sim 10^4$ - $10^5$.

\begin{figure}
\plottwo{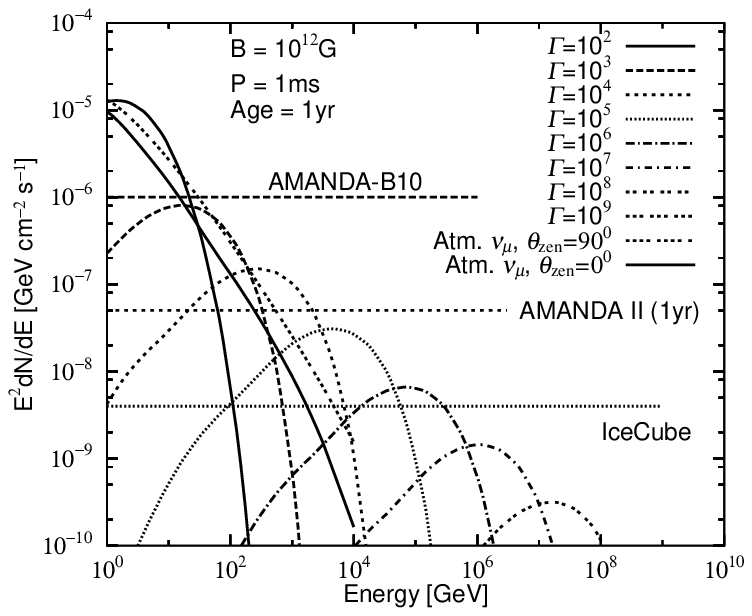}{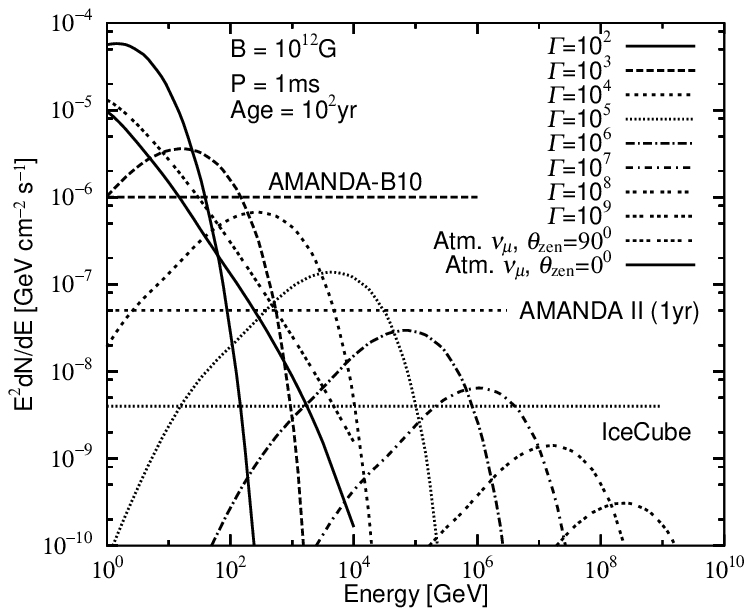}
\caption{
Spectrum of energy fluxes of neutrinos from a pulsar which is located 10 kpc
away from the earth. The amplitude of the magnetic field and period of
the pulsar is assumed to be $10^{12}$G and 1ms.
The minimum detectable energy flux of AMANDA-B10, AMANDA II (1yr),
and IceCube is represented by horizontal lines. The atmospheric
neutrino energy fluxes for a circular patch of  $1^{\circ}$ are also shown.
The left panel represents the case that
the age of the pulsar is 1yr (Models C2-C9),
while right panel represent the case that the age is 10$^2$yr (Models A2-A9). 
\label{fig9}}
\end{figure}

In figure~\ref{fig9}, we show spectrum of energy fluxes of neutrinos from
a pulsar which is located 10kpc away from the earth. The amplitude of the
magnetic field and period of the pulsar is assumed to be $10^{12}$G and 1ms.
The detection limits of the energy flux for AMANDA-B10, AMANDA II (1yr), and
IceCube are represented by horizontal lines. The atmospheric neutrino
energy fluxes for a circular patch of  $1^{\circ}$ are also shown~\cite{honda95}.
The left panel represents the case that the age of the pulsar is 1yr (Models C2-C9),
while right panel represent the case that the age is 10$^2$yr (Models A2-A9). 
From this figure, we can find that there is a possibility to detect the
signals of neutrinos from pulsar winds in our galaxy.  

\placefigure{fig9}

\begin{figure}
\plottwo{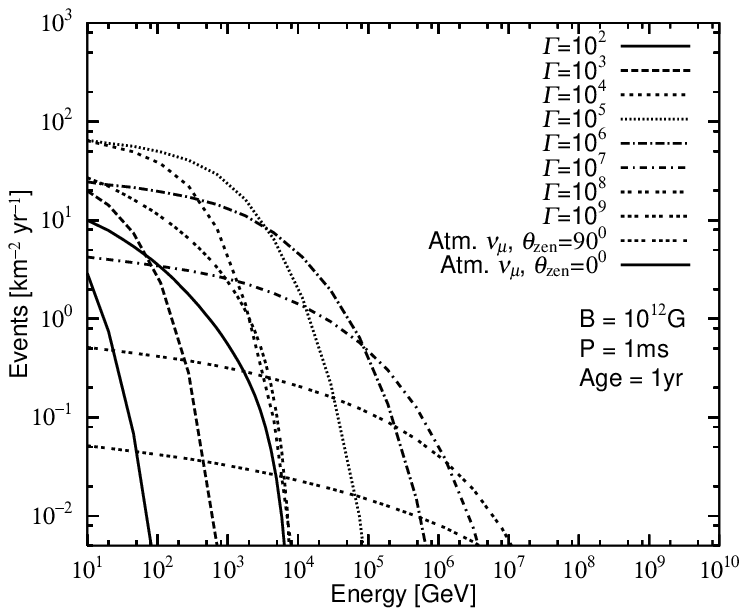}{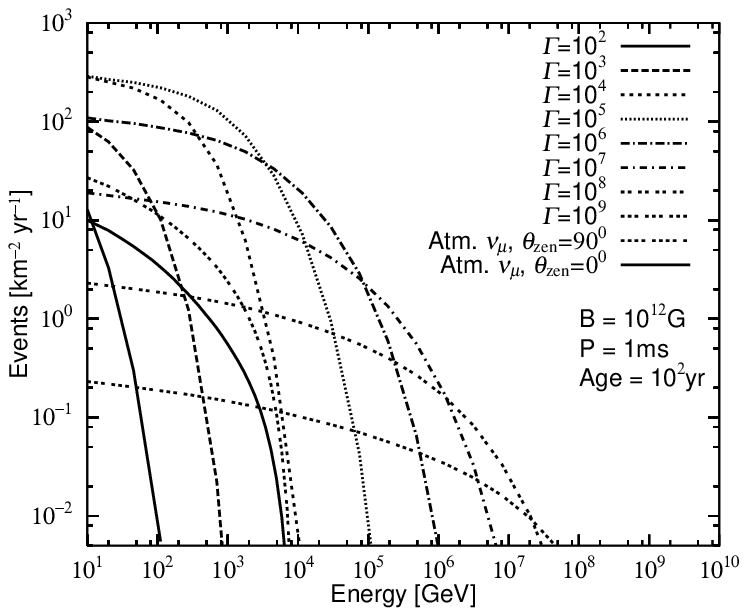}
\caption{
Neutrino event rate per year from a pulsar wind as a function of the muon
energy threshold in a km$^3$ high-energy neutrino detector. The pulsar is
assumed to be located 10 kpc away from the earth. The amplitude of the
magnetic field and period of the pulsar is assumed to be $10^{12}$G and 1ms.
The left panel represents the case that the age of the pulsar
is 1yr (Models C2-C9), while right panel represents the case that the age
is 10$^2$yr (Models A2-A9). The event rates expected of
atmospheric neutrino for a circular patch of  $1^{\circ}$ are also shown.
The event rates are shown
for the vertical downgoing atmospheric neutrinos and horizontal ones.  
\label{fig10}}
\end{figure}

In figure~\ref{fig10}, we show neutrino event rate per year from the nebula
flow considered in figure~\ref{fig9} as  a function of the muon energy
threshold in a km$^3$ high-energy neutrino detector. The left panel
represents the case that the age of the pulsar is 1yr (Models C2-C9), while right panel
represent the case that the age is 10$^2$yr (Models A2-A9). The event rates expected of
atmospheric neutrino for a circular patch of  $1^{\circ}$ are also shown~\cite{honda95}.
The event rates are shown for the vertical downgoing atmospheric neutrinos
and horizontal ones. It is confirmed that there is a possibility that the
signals from the nebula flow dominates the background of atmospheric
neutrinos. 

\placefigure{fig10}

\begin{figure}
\plottwo{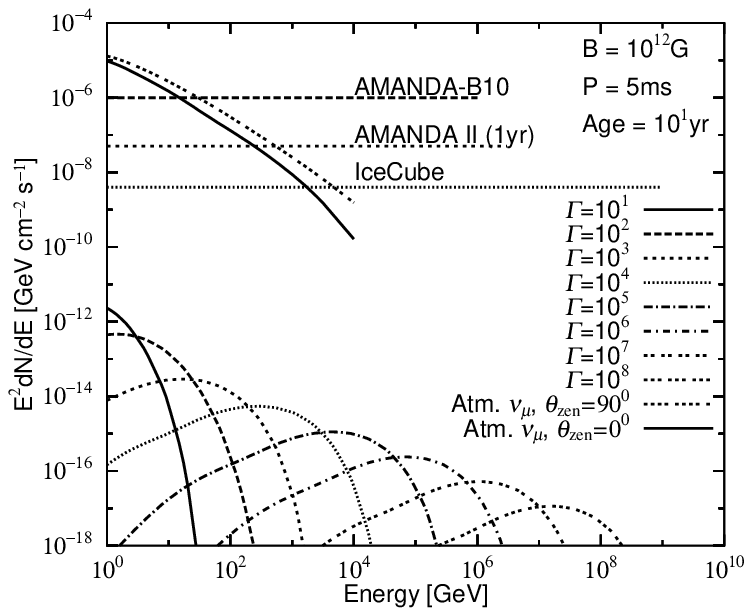}{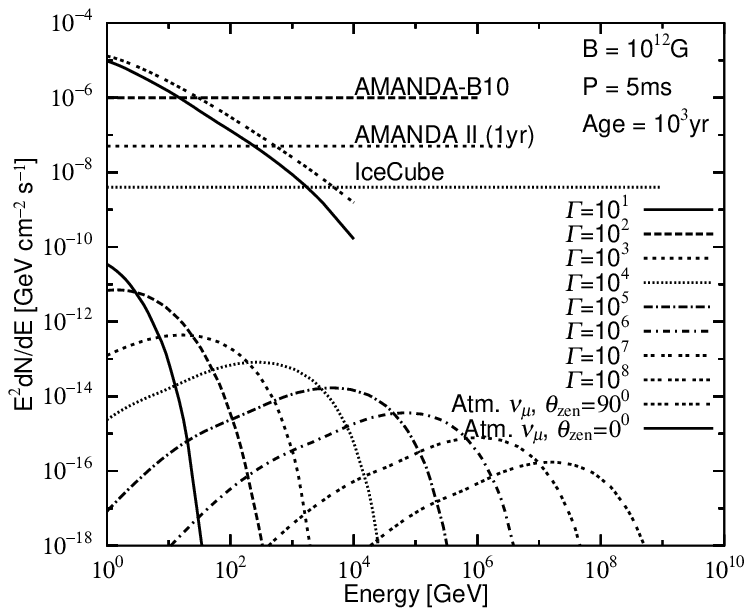}
\caption{
Same as figure 9, but for the period of the pulsar is 5ms.
The left panel represents the case that the age of the pulsar is 10yr
(Models G1-G8), while right panel
represent the case that the age is 10$^3$yr (Models E1-E8).
Note that the scale of the vertical line
is different from that of figure 9.  
\label{fig11}}
\end{figure}

For comparison, in figure~\ref{fig11}, we show the spectrum of energy fluxes of
neutrinos from a pulsar of which the
amplitude of the magnetic field and period are assumed to be
$10^{12}$G and 5ms. Location of the pulsar is assumed to be 10 kpc away
from the earth.
The left panel represents the case that the age of the
pulsar is 10yr (Models G1-G8), while right panel represents the case that the age is
10$^3$yr (Models E1-E8). In these cases, the fluxes of neutrinos are much lower than the
atmospheric neutrinos and detection limits of km$^3$ high-energy neutrino
detectors. We can conclude that the detectability of the signals from the
pulsar winds strongly depends on the period of the pulsar, which reflects
the luminosity of the pulsar winds (see Eq.(\ref{eqn25})).

\placefigure{fig11}

We check the reason why the flux of neutrinos becomes lower for the case
where period of the pulsar is 5ms. Figure~\ref{fig12} shows 
profiles of velocity, number density of protons, temperature, magnetic
field, and emissivity of charged pions in units of [10$^{-20}$erg s$^{-1}$
cm$^{-3}$] and [10$^{-20}$particles s$^{-1}$ cm$^{-3}$]. The left panel
represents the case that the period and age of the pulsar are 1ms and $10^2$
yr (Model A5; same with the right panel of figure 8), while the right panel shows the
case that the period and age of the pulsar are 5ms and $10^3$ yr (Model E5).
In both cases, initial bulk Lorenz factor of the pulsar wind is set to be
$10^5$.
In the
latter case, the location of the termination shock must be outward compared
with the former in order to achieve the pressure balance between the
outermost region of the nebula flow and innermost region of the remnant.
This is because luminosity of the pulsar wind and the pressure behind the
termination shock are lower in the latter case.
That is why the density of behind the termination shock is lower in the
latter case and results in lower emissivity of neutrinos.

\placefigure{fig12}

Here we consider the detectability of gamma-rays from neutral pions.
In figure~\ref{fig13}, integrated gamma-ray fluxes from the neutral pion
decays are shown assuming that the supernova ejecta has been optically thin
for gamma-rays~\cite{matz88}.
The amplitude of the magnetic field and period of the
pulsar are assumed to be $10^{12}$G and 1ms. The left panel represents
the case that the age of the pulsar is 1yr (Models C2-C9), while the right panel shows
the case that the age of the pulsar is $10^2$ yr (Models A2-A9).
The detection limits of integrated fluxes for GLAST, STACEE, CELESTE, HEGRA, CANGAROO,
MAGIC, VERITAS, and H.E.S.S.
are also shown. From these figures, we can find that there is a possibility
to detect gamma-rays from decays of neutral pions by these telescopes.
As for the spectrum of gamma-rays from electrons, it is difficult to
estimate it because we have to solve time evolution of energy distribution of
electrons taking into account the effects of cooling processes such as
synchrotron emission, inverse compton, and pair annihilation
as well as heating process such as Coulomb collisions between electrons and protons
and pair creations. This is difficult to solve and we regard it as a future work.

\placefigure{fig13}

On the other hand, we show the same figures in figure~\ref{fig14} 
but for the case that the period of the pulsar is 5ms.
The left panel represents the case that the age of the pulsar is 10yr (Models G1-G8),
while the right panel shows the case that the age of the pulsar is $10^3$ yr (Models E1-E8).
In these cases, the flux of gamma-rays is too low to detect.

\placefigure{fig14}

Finally, it is noted that there are some cases in which the location of the
termination shock is driven back to the surface of the neutron star since
the pressure balance can not be achieved (see section~\ref{boundary}).
Models H2 and H3 are the case (see table 2). In these cases, the density
behind the shock is so large that the resulting flux of neutrino is
very high. There will be a tendency that the reverse shock is likely to be
driven back to the surface of the neutron star when the age of the pulsar
is young, because the pressure at the innermost region of the remnant is
large (see figure~\ref{fig2}). Thus there may be a possibility to detect
a number of high energy neutrinos from the pulsar wind at the early stage
of the supernova explosion, although the protons in the nebula flow
will also suffer from synchrotron cooling and/or energy transfer to
electrons.

%%%%%%%%%%%%%%%%%%%%%%%%%%%%%%%%%%%%%%%%%%%%%%%%%%%%%%%%%%
\section{DISCUSSIONS}\label{discussions}
%%%%%%%%%%%%%%%%%%%%%%%%%%%%%%%%%%%%%%%%%%%%%%%%%%%%%%%%%%

\begin{figure}
\plottwo{fig8b.eps}{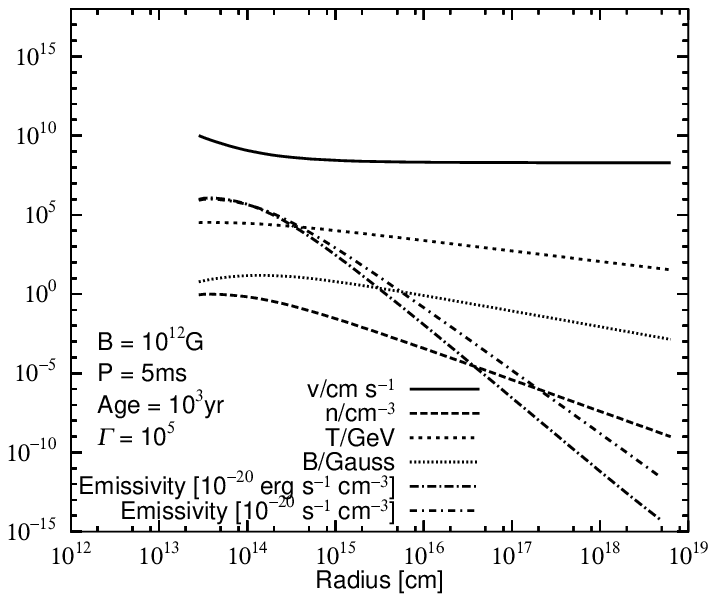}
\caption{
Profiles of velocity, number density of protons, temperature, magnetic field,
and emissivity of charged pions in units of
[10$^{-20}$erg s$^{-1}$ cm$^{-3}$] and
[10$^{-20}$particles s$^{-1}$ cm$^{-3}$]. The inner and outer boundaries
correspond to the location of the termination shock wave and the innermost
region of the surrounding supernova remnant.
The amplitude of the magnetic field is assumed to be $10^{12}$G.
Initial bulk Lorenz factor of the pulsar wind is set to be
$10^5$.
The left panel represents
the case that the period and age of the pulsar are 1ms and $10^2$ yr
(Model A5; same with the right panel
of figure 8), while the right panel
shows the case that the period and age of the pulsar are 5ms and
$10^3$ yr (Model E5).
\label{fig12}
}
\end{figure}

\begin{figure}
\plottwo{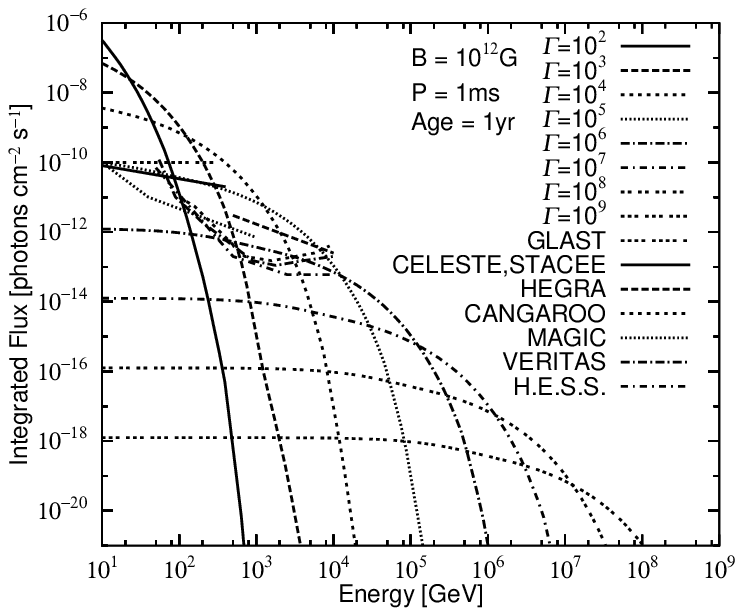}{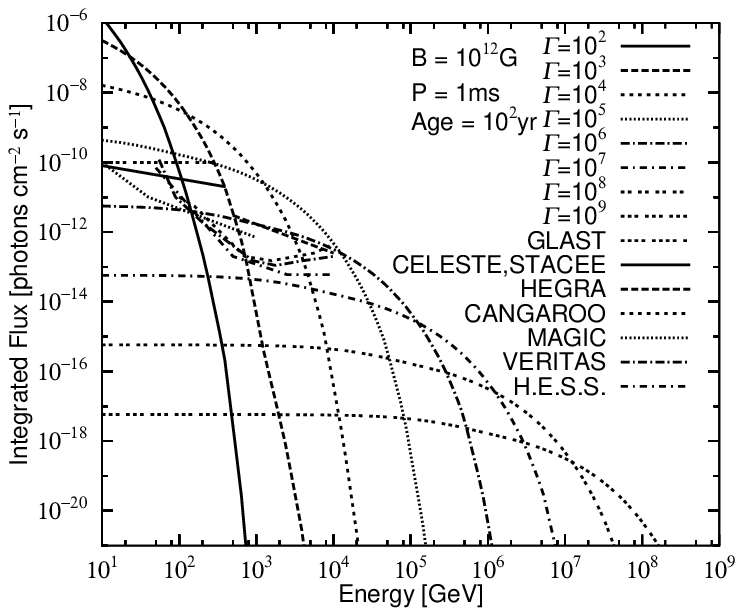}
\caption{
Integrated gamma-ray fluxes from the neutral pion decays are shown. 
The amplitude of the magnetic field and period of the pulsar are assumed to be
$10^{12}$G and 1ms. The left panel represents the case that the age of the
pulsar is 1yr (Models C2-C9), while the right panel shows the case that
the age of the pulsar is $10^2$ yr (Models A2-A9).
The minimum detectable integrated fluxes of GLAST, STACEE, CELESTE, HEGRA, CANGAROO,
MAGIC, VERITAS, and H.E.S.S.
are also shown.
\label{fig13}}
\end{figure}

\begin{figure}
\plottwo{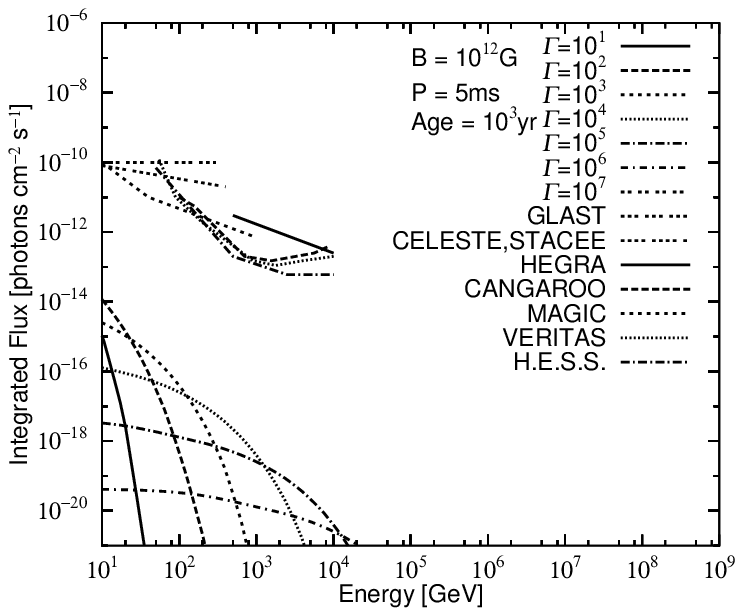}{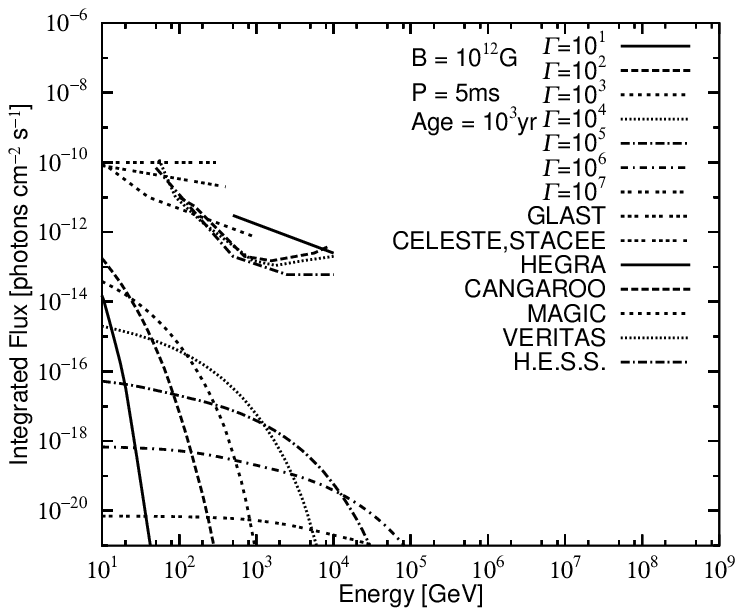}
\caption{
Same with figure 13, but for the case that the period of the pulsar is 5ms.
The left panel represents the case that the age of the pulsar is 10yr
(Models G1-G8),
while the right panel shows the case that the age of the pulsar is $10^3$
yr (Models E1-E8).
\label{fig14}}
\end{figure}

\begin{figure}
\plotone{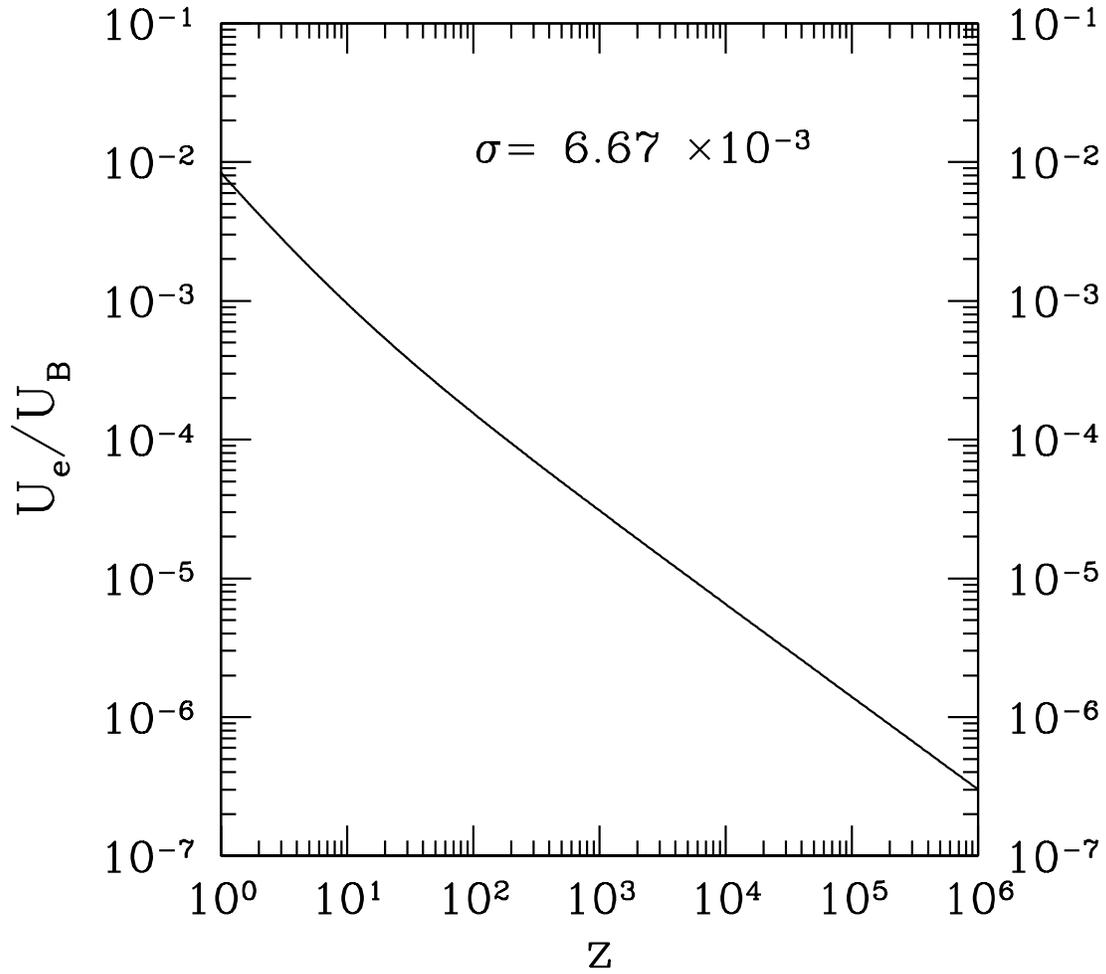}
\caption{
Ratio of the thermal energy of electrons relative to the energy of
electric-magnetic fields as a function of z ($\equiv r/r_s$) in the case where
$n_e$=$n_p$.
Definition of $\sigma$ is shown in Eq.(1) and set to
be $\sigma_c = 6.67 \times 10^{-3}$.
\label{fig15}}
\end{figure}

First, we discuss the timescale of energy loss of protons due to
the inverse compton (IC) scattering. This timescale depends on the
strength of the seed photons that suffers from great uncertainty.
However, we can estimate the lower limit of this timescale and show
that this process is not so effective. The ratio of the synchrotron
cooling timescale of protons relative to IC can be written as
\begin{eqnarray}
\frac{t_{\rm sync}}{t_{\rm IC}} = \frac{U_\gamma}{U_B},
\label{discussion-1}
\end{eqnarray}
where $U_\gamma$ and $U_B$ are energy density of seed photon and electric-magnetic
fields. In the case in which energy transfer from protons to electrons is ineffective,
the energy density of seed photon should be smaller than that of electrons
as long as the seed photons come from bremsstrahlung and/or synchrotron emission
from electrons.
Thus, using Eq.(\ref{eqnadd6}), the upper limit of the ratio of
$t_{\rm sync} /  t_{\rm IC}$ can be expressed as
\begin{eqnarray}
%\nonumber
\frac{t_{\rm sync}}{t_{\rm IC}} \le \frac{U_e}{U_B} = \frac{n_e m_e U_p}{n_p m_p U_B} =
\frac{n_e m_e}{n_p m_p}\frac{z^2\{3 \sigma + (3 \sigma/z)^{2/3}\}^2}{\sigma \{3 \sigma z^2 + (3 \sigma)^{2/3} z^{4/3}\}^{4/3} [4 + \{3 \sigma + (3 \sigma/z)^{2/3}\}^2]    }.
\label{discussion-2}
\end{eqnarray} 
We show in figure~\ref{fig15} the ratio of the thermal energy of electrons relative to the energy of
electric-magnetic fields as a function of z in the case where $n_e$=$n_p$. Here $\sigma$ is set to
be $\sigma_c = 6.67 \times 10^{-3}$. We can find that the upper limit of the ratio of 
$t_{\rm sync} /  t_{\rm IC}$ is smaller than unity as long as $n_e/n_p \lesssim 2 \times 10^{2}$.
That is why we can conclude that IC cooling is not so effective.
Here
we neglect the contribution of cosmic microwave-background, since it can be
expressed by an equivalent field strength $B_{\rm CMB} = 3.24 \mu$G, which is sufficiently
small and can be neglected. We also check the contribution of the photon field in the supernova
remnant around the nebula flow, since a portion of it may be transported into the region of
the nebula flow. The energy density of the photon field in the supernova remnant
can be estimated from figure~\ref{fig2} and expressed by
equivalent field strength $B_{\rm SN} \sim 3.9$ G, 1.3$\times 10^{-1}$ G,
4.1$\times 10^{-3}$ G, and 1.6$\times 10^{-4}$ G for $t = 1$, 10, 10$^2$, and 10$^3$ yr. These values
are also smaller than strength of the magnetic fields in almost all cases (see figures~\ref{fig8}
and~\ref{fig12}).
We should also estimate the number density of photons from surface of the pulsar.
The temperature of the surface of the neutrons star is considered to become in the range
($10^6 - 10^7$) K about 1 yr after the explosion, although there are some uncertainties
of the cooling processes~\cite{slane02}. Thus the energy density of the photon field at
radius $r$ can be estimated as
\begin{eqnarray}
U_{\gamma, \rm pulsar} = \frac{4 \pi R^2 \sigma T^4}{4 \pi r^2 c},
\label{discussion-2-2}
\end{eqnarray} 
% (1 + u^2) \left(   1 - \frac{u}{\sqrt{1 + u^2}}  \right)^2,
where $R$ is the radius of the pulsar and $\sigma$ is the Stefan-Boltzmann constant.
%Here we approximated that photons emitted from the surface of the pulsar are
% radiated away radially.
%Since $u \ll 1$, the factor $(1 + u^2) \left(   1 - u/\sqrt{1 + u^2}  \right)^2$
% is almost unity. Thus
% we regard this factor as unity for the further discussion.
Eq.~(\ref{discussion-2-2}) can be expressed by the equivalent field strength as
\begin{eqnarray}
B_{\rm pulsar} = 2.2 \times 10 \left(   \frac{R}{10^6 \rm cm}  \right) \left(   \frac{T}{10^7 \rm K}  \right)
\left(   \frac{10^{12} \rm cm}{r}  \right) \;\;\; \rm G,
\label{discussion-2-3}
\end{eqnarray} 
which is also smaller than strength of the magnetic fields in almost all cases (see figures~\ref{fig8}
and~\ref{fig12}).

\placefigure{fig15}

Next, we consider the effect of synchrotron cooling of muons and pions.
As for the mean life time of muon in the rest frame is 2.197$\times 10^{-6}$ s and 
that of charged pion is 2.603$\times 10^{-8}$ s.
%Thus, 
Since
synchrotron
cooling time of muon is
\begin{eqnarray}
\nonumber
t_{\mu, \rm syn} &=& \left( \frac{m_{\mu}}{m_e} \right)^4 \times
t_{e, \rm syn} \sim 1.82 \times 10^{9} t_{e, \rm syn} \\
                 &=& 7.11 \times 10^{10} \left( \frac{1 \; {\rm GeV}}{E_{\mu}} \right)
\left(  \frac{10^2 \; {\rm G}}{ B } \right)^2 \;\;\; \rm s,
%.
\label{discussion-3}
\end{eqnarray}
%Thus, 
the condition that the synchrotron cooling time becomes shorter than
that of the mean life time in the fluid rest frame can be expressed as
\begin{eqnarray}
E_{\mu} \ge 5.85 \times 10^5  \left(   \frac{10^4 \rm G}{B} \right) \; \rm GeV.
\label{discussion-4}
\end{eqnarray}
Since the amplitude of the magnetic field in the nebula flow becomes $\sim 10^4$ G at most
(see figures~\ref{fig8} and~\ref{fig12}), the energy spectrum of neutrinos whose energy is larger
than $\sim 10^5$ GeV may be modified due to the effect of the synchrotron cooling of muons.
Similarly, the energy spectrum of neutrinos whose energy is larger than $\sim 10^7$ GeV
may be modified by the effect of synchrotron cooling of charged pions.

Third, we discuss the cooling process due to photopion production which is interesting and should
be investigated in detail in the forth-coming paper because this is another process of producing
pions and may make the flux of neutrinos and gamma-rays enhance. Here we give the rough estimation
for this effect. The source of the seed photons are considered to be the synchrotron photons
emitted by electrons, the photon field in the supernova
remnant around the nebula flow, and photons from surface of the pulsar.
The typical frequency of the synchrotron photon is~\cite{rybicki79}
\begin{eqnarray}
\nu \sim 0.29 \times \frac{3}{8 \pi}  \gamma^2 \frac{e B}{m_e c} \sim 6.1 \times 10^5
\gamma^2  \left(\frac{B}{1 \rm G} \right), 
\label{discussion-5}
\end{eqnarray}
where $\gamma$ is the Lorenz factor of electrons in the fluid rest frame and is considered to be
almost same with the initial bulk Lorenz factor of the pulsar wind ($\equiv \Gamma$) as long as the electrons do not
lose their energy by synchrotron cooling. 
Here the average angle between the magnetic field and direction of motion of electron is set to be $\pi /4$.
This frequency corresponds to the energy
\begin{eqnarray}
E_{\nu} = 2.5 \times 10^{-9} \gamma^2  \left(\frac{B}{1 \rm G} \right) \; \rm eV.
\label{discussion-6}
\end{eqnarray}
Since there is a large peak in the photopion production cross section at $\epsilon \sim 0.35m_p c^2$
at the rest frame of a proton and the width of the peak cross section is also $\Delta \epsilon \sim 0.2$GeV~\cite{stecker79},
the energy of protons in the fluid rest frame which interact with the seed photons can be expressed as
\begin{eqnarray}
E_{p} \sim \frac{0.35m_p^2c^4}{E_{\nu}} = 1.2 \times 10^{13} \gamma^{-2} 
%\left(  \frac{10^4}{B} \right)
\left(  \frac{10^4 \rm G}{B} \right)
\; \rm GeV.
\label{discussion-7}
\end{eqnarray}
This shows, to be sure, that the protons in the nebula flow cause the photopion interactions
when the initial bulk Lorenz factor of the pulsar wind is larger than $\sim 10^5$. 
We roughly estimate the upper limit of the contribution of photopion production relative to
$pp$ interaction. The maximum number density of photons by synchrotron emission is estimated as
\begin{eqnarray}
\nonumber
n_{\gamma, \rm sync} \le \frac{U_e}{E_{\nu}} &=& \frac{n_e m_e}{n_p m_p} \frac{U_p}{E_{\nu}} \\
&\sim& \frac{m_e}{m_p} \frac{m_p c^2 \Gamma n_e}{E_{\nu}}
= 2.0 \times 10^5 
\left( \frac{10^5}{\Gamma} \right)  \left(  \frac{10^4 \rm G}{B}  \right) n_e.
\label{discussion-8}
\end{eqnarray}
Thus, if we assume that the energy loss rate per interaction is same between photopion production
and $pp$ interaction, and assume that the cross section of photopion production is constant ($\sim 5\times 10^{-28}$cm$^2$),
the contribution of photopion production relative to $pp$ interaction ($\sim 10^{-25}$cm$^2$) can be estimated to be
\begin{eqnarray}
\frac{\sigma_{p \gamma} n_{\gamma, \rm sync} }{\sigma_{pp} n_p} \sim 1.0 \times 10^3  \left( \frac{10^5}{\Gamma} \right)  \left(  \frac{10^4 \rm G}{B}  \right) \left(  \frac{n_e}{n_p}   \right)  .
\label{discussion-9}
\end{eqnarray}
In reality, as mentioned above,
there is a large peak in the photopion production cross section at the photon energy $\epsilon = 0.35m_pc^2$
in the proton rest frame due to the $\Delta(1232)$ resonance~\citep{stecker79,waxman97,amato03}.
Thus the contribution of photopion production would be much
smaller than the estimated value in Eq.(\ref{discussion-9}) in the case that the typical energy of protons which interact
with the synchrotron photons from electrons through the $\Delta(1232)$ resonance is much different from the mean
energy of protons in the fluid. We emphasize again that we estimated here just the upper limit of the contribution
of photopion production. Also, it is noted here that the synchrotron cooling time of electrons can be estimated as
\begin{eqnarray}
t_{e, \rm sync} = 7.6 \times 10^{-5} \left(   \frac{10^5}{\Gamma}  \right) \left(  \frac{10^4 \rm G}{B}  \right)^2 \; \rm s.
\label{discussion-10}
\end{eqnarray}
Thus, even if the effect of the photopion production can not be neglected, it will be effective only within a very short
distance from the location of the termination shock. As for the contribution of the photon field from the supernova
remnant, the energy of protons that interacts with the photon field can be estimated as
\begin{eqnarray}
E_p \sim \frac{0.35 m_p^2c^4}{kT_{\gamma}} \sim 3.1 \times 10^9 \left( \frac{0.1 \rm eV}{kT_{\gamma}} \right) \; \rm GeV,
\label{discussion-11}
\end{eqnarray}
where $T_{\gamma}$  is the temperature of the photon field. Thus the required energy for a proton to interact
with the photon field from the supernova remnant seems to be too high.
As for the contribution of the photon field from the pulsar, the number density of the photon field can be expressed as
\begin{eqnarray}
\nonumber
n_{\gamma,\rm pulsar} &=& \frac{4 \pi R^2 \sigma T^4}{4 \pi r^2 c} \left(   \frac{1}{k T}   \right) \\
&=& 1.4 \times 10^{10} \left(  \frac{10^{12} \rm cm}{r}  \right)^2  \left(  \frac{R}{10^{6} \rm cm}  \right)^2
\left(  \frac{T}{10^{7} \rm K}  \right)^3 \; \rm cm^{-3}.
\label{discussion-11-2}
\end{eqnarray}
Thus, from Eqs.(\ref{discussion-8}) and (\ref{discussion-11-2}), the photon fields from the pulsar
can dominate the synchrotron photons. As for the energy of protons that interacts with the photon field,
it can be estimated as
\begin{eqnarray}
E_p \sim 3.6 \times 10^5 \left( \frac{10^7 \rm K}{T} \right) \; \rm GeV.
\label{discussion-11-3}
\end{eqnarray}
Thus, protons with energies $\sim 10^5$ GeV may interact with the photons from the
pulsar and produce much more pions than estimated in this study. Thus, the spectrum
of neutrinos and gamma-rays around $\sim 10^5$ GeV may be modified and intensity may be
enhanced due to this effect. This effect is very important and should be investigated
carefully in the very near future.

In this study, we consider the pion production due to $pp$ interaction. However, in many cases,
total energy of colliding protons in their center-of-mass system exceeds the experimental limit.
Moreover, it exceeds the mass of the top quark ($m_t = 174.3 \pm 5.1$GeV from direct observation of top
events, and $m_t = 178.1^{+ 10.4}_{-8.3}$GeV from standard model electroweak fit~\cite{hagiwara02}). 
As explained in section~\ref{cross}, it is confirmed that the scaling
assumption does not much affect the resulting gamma-ray spectrum by
comparing the resulting gamma-ray spectrum up to $\sim 10^7$ GeV with the gamma-ray production
by the mini-jet model including the QCD effect even if the scaling
model is extrapolate to higher energies. However, it should be confirmed whether the scaling
assumption is still valid at more high energies, which is investigated in the near future.

In this study, we have estimated the energy transfer timescale assuming that it is mainly
caused by Coulomb scattering between protons and electrons~\cite{stepney83}. 
In this case, the energy transfer timescale depends very weakly on the number ratio 
of protons to electrons (positrons), as shown in section~\ref{results}. However,
it is pointed out by Hoshino et al. (1992) that the efficiency of the energy transfer from protons
to positrons may be higher than that from protons to electrons due to the effect of cyclotron
resonance. Thus, if this process is really effective and determines the energy transfer rate,
number ratio of protons to electrons (positrons) should be important because the energy transfer
timescale will strongly depend on the number ratio. Also, the expected flux of neutrinos
and gamma-rays from $pp$ interactions will be decreased than estimated in this study. 
In this case, however, emission of gamma-rays from electron-positron plasma should be
important through the process of synchrotron, inverse compton, and pair annihilation.
Neutrino emission from electron-positron plasma is also expected from electron-positron
pair annihilation.
In more realistic cases, there will be heavy nuclei in the pulsar wind, which should be
taken into account as a next step of this study. In this case, hadronic interactions
should be enhanced since heavy nuclei are composed of protons and neutrons and baryon number should be
enhanced. In order to obtain the contribution of heavy nuclei, photodisintegration of heavy nuclei should be
also taken into account, which will make the situation rather complicated.

In this study, we assumed that the pressure balance is hold at the interface between
the nebula flow and supernova remnant. Also, we fixed the fraction of magnetic field energy
in the pulsar wind so that the velocity of the nebula flow is almost same with that of the
supernova remnant at the interface, which makes the location of the termination shock fixed.
However, in reality, the pressure balance should not be always hold and velocity of the
nebula flow and supernova remnant will be different in general. This means that we need to
consider the momentum transfer between the nebula flow and supernova remnant, which will change
profiles of density, temperature, and amplitude of the magnetic field in the nebula flow.
This is a future work and it may require numerical calculations. Also, it is assumed that
the system is spherical symmetry for simplicity. Of course, in reality, we should consider
the effects of geometrical anisotropy of the system~\cite{hester02}.

In this study, we investigated the effects that bulk Lorenz factor is dissipated into
the random motion of the charged particles when they pass through the termination shock.
Of course, a part of charged particles will be accelerated at more high energies due to
Fermi acceleration~\cite{blandford78}. Maximum energy of protons due to Fermi acceleration can be estimated
using Larmor radius as
\begin{eqnarray}
E_{\rm max} = 3.3 \times 10^9 \left( \frac{B}{10^4 \rm G}  \right) \left( \frac{r}{10^{12} \rm cm}  \right) \; \rm GeV. 
\label{discussion-12}
\end{eqnarray}
From figures~\ref{fig8} and \ref{fig12}, $E_{\rm max}$ will be about 10$^9$ GeV and 10$^8$ GeV for the cases of
$P=1$ms and 5ms ($B_p$ is set to be 10$^{12}$G). Thus, the flux and mean energies of neutrinos
and gamma-rays may be enhanced due to the effects of Fermi acceleration. We will have to investigate
carefully how much of charged particles are accelerated by Fermi acceleration in order to
estimate this effects quantitatively.

It is noted that the $\nu_{\tau}$ is generated due to the effect of neutrino oscillation~\cite{fukuda98}.
The scale length for $\nu_{\mu}$ with energy $E_{\nu}$ to be converted to be $\nu_{\tau}$ due to the vacuum oscillation
is written as
\begin{eqnarray}
L = 4.9 \times 10^{10} \left( \frac{E_{\nu}}{10^3 \rm GeV}   \right) \left(    \frac{2.5\times 10^{-3} \rm eV^2}{\Delta m^2}   \right) \; \rm cm,
\label{discussion-13}
\end{eqnarray}
where $\Delta m$ is the mass difference between $m_2$ and $m_3$.
Thus the neutrinos produced in the nebula flow are well mixed and ratios of the flux among three
flavors ($e$, $\mu$, and $\tau$) will become order unity at the earth. 
Thus there may be a possibility to detect the double-bang events (a big hadronic shower from the initial
$\nu_{\tau}$ interaction and a second big particle cascade due to $\tau$ decay, which was pointed out
by Learned and Pakvasa 1995)
at the km$^3$ neutrino telescopes from the system investigated in this study.

At present, there is no candidate of a pulsar whose wind is more luminous than $\sim 10^{43}$
% erg
erg s$^{-1}$
like the case of $B=10^{12}$G and $P=1$ms in our galaxy~\citep{manchester99,torres03}. This is
consistent with the fact that there is no point-like source detected at AMANDA-B10 detector~\cite{ahrens03}.
We hope that such point-like sources are found at AMANDA-II and/or IceCube detectors in the
near future. Also, we hope that such an active pulsar is found by pulsar surveys in
radio, X-ray, and gamma-ray bands. We also hope that a newly born supernova will appear in our
galaxy in the near future.

Recently, Granot and Guetta (2003) pointed out the possibility to detect neutrinos from pulsar nebula
that is associated with a gamma-ray burst (GRB). It will be interesting to apply our model to GRB
scenario and estimate fluxes of neutrino background by setting the amplitude of magnetic field
at the pole of the pulsar is set to be ($10^{14}$ - $10^{15}$) G~\cite{granot03}.
Information on GRB formation rate history and explosion mechanism of GRB
may be obtained if such signals are detected~\citep{nagataki03,nagataki03b}.

%%%%%%%%%%%%%%%%%%%%%%%%%%%%%%%%%%%%%%%%%%%%%%%%%%%%%%%%%%
\section{SUMMARY AND CONCLUSION}\label{conclusion}
%%%%%%%%%%%%%%%%%%%%%%%%%%%%%%%%%%%%%%%%%%%%%%%%%%%%%%%%%%

In this study, we have estimated fluxes of neutrinos and gamma-rays that are
generated from decays of charged and neutral pions from a
pulsar surrounded by supernova ejecta in our galaxy, including
an effect that has not been taken into consideration, that is, interactions
between high energy cosmic rays themselves in the nebula flow, assuming that
hadronic components be the energetically dominant species in the pulsar wind.
Bulk flow is assumed to be randomized by passing through the termination shock
and energy distribution functions of protons and electrons behind the termination
shock obey the relativistic Maxwellians.

We have found that fluxes of neutrinos and gamma-rays depend very sensitively on the
wind luminosity, which is assumed to be comparable with the spin-down luminosity.
In the case where $B=10^{12}$G and $P=1$ms, neutrinos should be detected by
km$^3$ high-energy neutrino detectors such as AMANDA and IceCube. Also, gamma-rays
should be detected by Cherenkov telescopes such as CANGAROO and H.E.S.S.
as well as
by gamma-ray satellites such as GLAST. On the other hand, in the case where $B=10^{12}$G
and $P=5$ms, fluxes of neutrinos will be too low to be detected even
by the next-generation detectors. However, even in the case where $B=10^{12}$G
and $P=5$ms, there is a possibility that very high flux of neutrinos and gamma-rays
may be realized at the early stage of a supernova explosion ($t \le 1$yr), where the
location of the termination shock is very near to the pulsar (ex. Model H3 and H4 in table 2).  
We also found that there is a possibility that protons with energies $\sim 10^5$ GeV
in the nebula flow may interact with the photon field from surface of the
pulsar and produce much
pions, which enhances the intensity of resulting neutrinos and gamma-rays.

We have found that interactions between high energy cosmic rays themselves
are so effective that this effect can be confirmed by future observations.
Thus, we conclude that it is worth while investigating this effect further
in the near future.

\acknowledgments
The author thanks K. Sato, M. Hoshino, S. Shibata, S. Yamada, S. Yoshida, T. Naito, and K. Kohri for useful comments.
The author is also pleased to acknowledge useful comments by the anonymous referee of this manuscript. 
This work is supported in part by Grants-in-Aid for scientific research provided by
the Ministry of Education, Science and Culture of Japan through Research Grant
No.S 14102004 and No. 14079202.

\appendix
%%%%%%%%%%%%%%%%%%%%%%%%%%%%%%%%%%%%%%%%%%%%%%%%%%%%%%%%%%
\section{Estimation of the Adiabatic Index}\label{appa}
%%%%%%%%%%%%%%%%%%%%%%%%%%%%%%%%%%%%%%%%%%%%%%%%%%%%%%%%%%
According to Hoshino et al. (1992), the distribution functions of protons
behind the termination shock are almost exactly described by relativistic
Maxwellians as
\begin{eqnarray}
N(\gamma) = A \gamma \exp \left[   -\frac{m_pc^2}{k_B T} (\gamma -1)   \right],
\label{eqna1}
\end{eqnarray}
where $k_B$ is the Boltzmann constant and $A$ is the normalization factor
in units of cm$^{-3}$. In this case, the number density ($n$) [cm$^{-3}$],
energy density ($e$) [erg cm$^{-3}$], and pressure ($P$) [dyn cm$^{-2}$] can
be written as
\begin{eqnarray}
n = A e^{\alpha} \int_1^{\infty} \gamma e^{-\alpha \gamma} d \gamma
= A \left[  \frac{1}{\alpha} + \frac{1}{\alpha^2} \right],
\label{eqna2}
\end{eqnarray} 
\begin{eqnarray}
\nonumber
e &=& \int E  \left( \frac{dn}{d^3xd^3p} \right) d^3p = \int m_pc^2\gamma
\frac{N(\gamma)}{4 \pi p^2 \frac{dp}{d\gamma}} 4\pi p^2 \frac{m_pc\gamma}{\sqrt{\gamma^2-1}} d\gamma
=
Am_p c^2 e^{\alpha} \int_1^{\infty}   \gamma^2 e^{-\alpha \gamma}
d \gamma \\ 
&=& Ak_B T  \left[ 1 + \frac{2}{\alpha} + \frac{2}{\alpha^2} \right],
\label{eqna3}
\end{eqnarray} 
and
\begin{eqnarray}
\nonumber
P &=& \frac{1}{3} \int pv \left( \frac{dn}{d^3xd^3p} \right) d^3p
= \frac{1}{3} \int p \frac{pc^2}{E} \frac{N(\gamma)}{4 \pi p^2 
\frac{dp}{d\gamma}} 4 \pi p^2 dp
= \frac{Am_pc^2}{3}e^\alpha \int_1^{\infty} (\gamma^2 -1) e^{- \frac{m_pc^2}{k_BT}\gamma} d\gamma \\
&=& \frac{2}{3}Ak_B T  \left[ \frac{1}{\alpha} + \frac{1}{\alpha^2} \right],
\label{eqna4}
\end{eqnarray} 
where $\alpha = m_pc^2/k_BT$, $E=m_pc^2\gamma$, $dn/d^3xd^3p$ is the number
density in phase space. In this case, the relation $P = 2 n k_B T/3$ holds
exactly, and in the high temperature limit ($\alpha \ll 1$), the relation
\begin{eqnarray}
E = eV = 3PV
\label{eqna5}
\end{eqnarray} 
holds approximately, where $V$ is the volume [cm$^3$]. 
In this case, the relation $3VdP = -4P dV$ holds in the adiabatic
transition. Thus $P \propto V^{-4/3} \propto \rho^{4/3}$ and adiabatic
index becomes 4/3 in high temperature limit.

%%%%%%%%%%%%%%%%%%%%%%%%%%%%%%%%%%%%%%%%%%%%%%%%%%%%%%%%%%%%%%%%%%%%%%%%%%
\section{Derivation of Number Flux of Pions in Observer's Frame}\label{appb}
%%%%%%%%%%%%%%%%%%%%%%%%%%%%%%%%%%%%%%%%%%%%%%%%%%%%%%%%%%%%%%%%%%%%%%%%%%

We consider an fluid element and number spectrum of pions produced
in the fluid element. Using the number spectrum of pions per solid angle
[particles s$^{-1}$ erg$^{-1}$ sr$^{-1}$], the conservation law of number of
particles can be expressed as
\begin{eqnarray}
\int dE_{\nu} \int \sin \theta d \theta d \phi \frac{1}{4 \pi} F(E_{\nu}) dt
= \int d \bar{E}_{\nu} \int \sin \bar{\theta} d \bar{\theta} d \bar{\phi} \frac{d}{d \Omega} \bar{F}(\bar{E}_{\nu})
dt \bar{\Gamma} (1- \bar{\beta}  \cos \bar{\theta}),
\label{eqnb1}
\end{eqnarray} 
where bars are labeled for the quantum of the observer's frame,
while the quantum without label denote the ones in the fluid-rest frame.
The axis $\theta$ = $\bar{\theta}$ = 0 is set to be aligned with the
direction of the observer (i.e. the earth) measured from the fluid element.
$\bar{\Gamma}$ is the bulk Lorenz factor of the fluid element in the observer's
frame. It is noted that $dt_{A} = \bar{\Gamma} (1- \bar{\beta} ) dt$ is the time
interval of the radiation as received by a stationary receiver in the
observer's frame~\cite{rybicki79}.
Here we define the quantum $N(\theta)$ as
\begin{eqnarray}
N(\theta) = \int dE_{\nu} F(E_{\nu}) \;\; \rm [particles \; s^{-1}].
\label{eqnb2}
\end{eqnarray} 
Using this expression, the relation
\begin{eqnarray}
\frac{d}{d \bar{\Omega}} \bar{N}( \bar{\theta}) = \frac{1}{\bar{\Gamma} (1- \bar{\beta}  \cos \bar{\theta})}
\frac{\sin \theta}{\sin \bar{\theta}} \frac{d \phi}{d \bar{\phi}}
\frac{1}{4 \pi} N(\theta)
\label{eqnb3}
\end{eqnarray} 
can be derived. Using the relation $\cos \bar{\theta}$ = $(\cos \theta + \beta)
/(1+\beta \cos \theta)$ and $d \phi = d \bar{\phi}$, Eq.~(\ref{eqnb3}) can be
expressed as
\begin{eqnarray}
\frac{d}{d \bar{\Omega}} \bar{N}(\bar{\theta}) = \frac{1}{\bar{\Gamma}^{3}}
\frac{1}{(1 - \bar{\beta} \cos \bar{\theta})^3} \frac{1}{4 \pi}N(\theta). 
\label{eqnb4}
\end{eqnarray} 
Also, the relation $\bar{F}(\bar{E}_{\nu})  = F(E_{\nu})dE_{\nu}/d\bar{E}_{\nu}$
can be expressed as
\begin{eqnarray}
\bar{F}(\bar{E}_{\nu}) = \frac{1}{\bar{\Gamma}(1+ \bar{\beta} \cos \theta)} F(E_{\nu}) = \bar{\Gamma}(1 - \bar{\beta}
 \cos \bar{\theta}) F(E_{\nu}).
\label{eqnb5}
\end{eqnarray}
Finally, using Eqs.~(\ref{eqnb4}) and (\ref{eqnb5}), the relation
\begin{eqnarray}
\frac{d}{d \bar{\Omega}} \bar{F}(\bar{E}_{\nu}) = \frac{1}{\bar{\Gamma^2}(1- \bar{\beta} \cos \bar{\theta})^2}
\frac{1}{4 \pi} F(E_{\nu})
\label{eqnb6}
\end{eqnarray} 
is derived.

%%%%%%%%%%%%%%%%%%%%%%%%%%%%%%%%%%%%%%%%%%%%%%%%%%%%%%%%%%%%%%%%%%%%%%%%%%%%%%
\section{Analytical Solution for the Nebula Flow}\label{appc}
%%%%%%%%%%%%%%%%%%%%%%%%%%%%%%%%%%%%%%%%%%%%%%%%%%%%%%%%%%%%%%%%%%%%%%%%%%%%%%

The analytical solution of equations for the nebula flow is obtained by
Kennel and Coroniti (1984), which is expressed as
\begin{eqnarray}
(1+ u_2^2v^2)^{1/2} \left[  \delta + \Delta (vz^2)^{-1/3} + \frac{1}{v}
       \right] = \Gamma_2 (1+\delta+\Delta),
\label{eqnadd1}
\end{eqnarray}
where $v$ is defined as $v = u/u_2$, $z$ is defined as $z=r/r_s$,
$\delta$ is defined as $\delta= 4 \pi n_2 \Gamma_2^2 m_pc^2/B_2^2 \sim
u_2/u_1\sigma$, and $\Delta$ is defined as $16 \pi P_2 \Gamma_2^2/B_2^2$. 
The solution for this equation can be written as
\begin{eqnarray}
v = \frac{1}{2(1+ \Delta)}\frac{G^3}{x^2},
\label{eqnadd2}
\end{eqnarray}
where
\begin{eqnarray}
%\nonumber
G = 1 + \{ 1 + x^2 + \left[  (1+x^2)^2 -1         \right]^{1/2} \}^{1/3}
      + \{ 1 + x^2 - \left[  (1+x^2)^2 -1         \right]^{1/2} \}^{1/3},
\label{eqnadd3}
\end{eqnarray}
\begin{eqnarray}
x = z/z_{\Delta},
\label{eqnadd4}
\end{eqnarray}
and
\begin{eqnarray}
z_\Delta = \left[   \frac{2}{27} \frac{\Delta^3}{(1+\Delta)^2}  \right]^{1/2}.
\label{eqnadd5}
\end{eqnarray}

Using this solution, we can derive the solution for the total pressure
$P_{\rm T}$ in the postshock in Eq.~(\ref{eqnadd6}).

\clearpage


\clearpage

\begin{deluxetable}{cccccccccccccc}
%\tabletypesize{\scriptsize}
\tablecaption{Parameters for the representation of invariant cross section}
\tablewidth{0pt}
\tablehead{
\colhead{Particle}      & \colhead{$A$}           & \colhead{$B$}   &
\colhead{$r$}           & \colhead{$C_1$}         & \colhead{$C_2$} & 
\colhead{$C_3$} \\
                        & [mb/(GeV$^2$/c$^3$)]    & [(GeV/c)$^{-1}$]& 
                        &                         & [(GeV/c)$^{-1}$]&
[(GeV/c)$^{-2}$]& 
}
\startdata
$\pi^{0}$               & 140                      &5.43             &
2                       &6.1                       &-3.3             &
0.6\\
$\pi^{+}$               & 153                      &5.55             &
1                       &5.37                      &-3.5             &
0.83\\
$\pi^{-}$               & 127                      &5.30             &
3                       &7.03                      &-4.5             &
1.67\\
 \enddata
\label{tab1}
\end{deluxetable}

\begin{deluxetable}{llllllccclllllllll}
\tabletypesize{\scriptsize}
\tablecaption{Input and Output Parameters}
\tablewidth{0pt}
\tablehead{
\colhead{Model}             &
\colhead{$B$}               & \colhead{$P$}           & \colhead{$L$}         &  \colhead{Age}              & \colhead{$\Gamma$}    &        
\colhead{$r_{\rm shock}$}   & \colhead{$t_{\rm min}$\tablenotemark{a}}        & \colhead{min($t_{\rm min}$,$t_{\rm ep}$)\tablenotemark{b}}                                 & \colhead{$L_{\pi}$}& 
\colhead{Event Rate}\\
                            &
 \colhead{[G]}              & [ms]                    & [erg s$^{-1}$]        &   [yr]                      &                       &  
 [cm]                       &                         &                       &   [erg s$^{-1}$]            &     
[km$^{-2}$ yr$^{-1}$]} 
\startdata
Model A1                    &
$10^{12}$                   & 1                       & $9.6\times 10^{42}$   &  $10^{2}$                   & $10^1$                &  
$1.5\times 10^{12}$         & $t_{\rm travel}$        & $t_{\rm ep}$          &  ------------               & ------------            \\
Model A2                    &
$10^{12}$                   & 1                       & $9.6\times 10^{42}$   &  $10^{2}$                   & $10^2$                &  
$1.5\times 10^{12}$         & $t_{\rm travel}$        & $t_{\rm travel}$      &  $5.5\times 10^{39}$        & $1.3\times 10^{1}$       \\
Model A3                    &
$10^{12}$                   & 1                       & $9.6\times 10^{42}$   &  $10^{2}$                   & $10^3$                &  
$1.5\times 10^{12}$         & $t_{\rm travel}$        & $t_{\rm travel}$      &  $2.6\times 10^{38}$        & $8.8\times 10^{1}$      \\
Model A4                    &
$10^{12}$                   & 1                       & $9.6\times 10^{42}$   &  $10^{2}$                   & $10^4$                &  
$1.5\times 10^{12}$         & $t_{\rm travel}$        & $t_{\rm travel}$      &  $2.6\times 10^{37}$        & $2.8\times 10^{2}$      \\
Model A5                    &
$10^{12}$                   & 1                       & $9.6\times 10^{42}$   &  $10^{2}$                   & $10^5$                &  
$1.5\times 10^{12}$         & $t_{\rm travel}$        & $t_{\rm travel}$      &  $2.7\times 10^{36}$        & $2.9\times 10^{2}$       \\
Model A6                    &
$10^{12}$                   & 1                       & $9.6\times 10^{42}$   &  $10^{2}$                   & $10^6$                &  
$1.5\times 10^{12}$         & $t_{\rm travel}$        & $t_{\rm travel}$      &  $2.7\times 10^{35}$        & $1.1\times 10^{2}$       \\
Model A7                    &
$10^{12}$                   & 1                       & $9.6\times 10^{42}$   &  $10^{2}$                   & $10^7$                &  
$1.5\times 10^{12}$         & $t_{\rm travel}$        & $t_{\rm travel}$      &  $2.7\times 10^{34}$        & $1.9\times 10^{1}$      \\
Model A8                    &
$10^{12}$                   & 1                       & $9.6\times 10^{42}$   &  $10^{2}$                   & $10^8$                &  
$1.5\times 10^{12}$         & $t_{\rm travel}$        & $t_{\rm travel}$      &  $2.7\times 10^{33}$        & $2.3\times 10^{0}$      \\
Model A9                    &
$10^{12}$                   & 1                       & $9.6\times 10^{42}$   &  $10^{2}$                   & $10^9$                &  
$1.5\times 10^{12}$         & $t_{\rm travel}$        & $t_{\rm travel}$      &  $2.7\times 10^{32}$        & $2.3\times 10^{-1}$    \\
Model B1                    &
$10^{12}$                   & 1                       & $9.6\times 10^{42}$   &  $10^{1}$                   & $10^1$                &  
$4.3\times 10^{12}$         & $t_{\rm travel}$        & $t_{\rm ep}$          &  ------------               &  ------------           \\
Model B2                    &
$10^{12}$                   & 1                       & $9.6\times 10^{42}$   &  $10^{1}$                   & $10^2$                &  
$4.3\times 10^{12}$         & $t_{\rm travel}$        & $t_{\rm travel}$      &  $1.5\times 10^{39}$        & $3.6\times 10^{0}$       \\
Model B3                    &
$10^{12}$                   & 1                       & $9.6\times 10^{42}$   &  $10^{1}$                   & $10^3$                &  
$4.3\times 10^{12}$         & $t_{\rm travel}$        & $t_{\rm travel}$      &  $7.5\times 10^{37}$        & $2.5\times 10^{1}$      \\
Model B4                    &
$10^{12}$                   & 1                       & $9.6\times 10^{42}$   &  $10^{1}$                   & $10^4$                &  
$4.3\times 10^{12}$         & $t_{\rm travel}$        & $t_{\rm travel}$      &  $7.5\times 10^{36}$        & $8.0\times 10^{1}$      \\
Model B5                    &
$10^{12}$                   & 1                       & $9.6\times 10^{42}$   &  $10^{1}$                   & $10^5$                &  
$4.3\times 10^{12}$         & $t_{\rm travel}$        & $t_{\rm travel}$      &  $7.6\times 10^{35}$        & $8.2\times 10^{1}$      \\
Model B6                    &
$10^{12}$                   & 1                       & $9.6\times 10^{42}$   &  $10^{1}$                   & $10^6$                &  
$4.3\times 10^{12}$         & $t_{\rm travel}$        & $t_{\rm travel}$      &  $7.6\times 10^{34}$        & $3.1\times 10^{1}$      \\
Model B7                    &
$10^{12}$                   & 1                       & $9.6\times 10^{42}$   &  $10^{1}$                   & $10^7$                &  
$4.3\times 10^{12}$         & $t_{\rm travel}$        & $t_{\rm travel}$      &  $7.6\times 10^{33}$        & $5.3\times 10^{0}$       \\
Model B8                    &
$10^{12}$                   & 1                       & $9.6\times 10^{42}$   &  $10^{1}$                   & $10^8$                &  
$4.3\times 10^{12}$         & $t_{\rm travel}$        & $t_{\rm travel}$      &  $7.6\times 10^{32}$        & $6.5\times 10^{-1}$   \\
Model B9                    &
$10^{12}$                   & 1                       & $9.6\times 10^{42}$   &  $10^{1}$                   & $10^9$                &  
$4.3\times 10^{12}$         & $t_{\rm travel}$        & $t_{\rm travel}$      &  $7.6\times 10^{31}$        & $6.5\times 10^{-2}$     \\
Model C1                    &
$10^{12}$                   & 1                       & $9.6\times 10^{42}$   &  $10^{0}$                   & $10^1$                &  
$4.8\times 10^{12}$         & $t_{\rm travel}$        & $t_{\rm ep}$          &  ------------               & ------------            \\
Model C2                    &
$10^{12}$                   & 1                       & $9.6\times 10^{42}$   &  $10^{0}$                   & $10^2$                &  
$4.8\times 10^{12}$         & $t_{\rm travel}$        & $t_{\rm travel}$      &  $5.9\times 10^{38}$        & $2.9\times 10^{0}$     \\
Model C3                    &
$10^{12}$                   & 1                       & $9.6\times 10^{42}$   &  $10^{0}$                   & $10^3$                &  
$4.8\times 10^{12}$         & $t_{\rm travel}$        & $t_{\rm travel}$      &  $5.9\times 10^{37}$        & $2.0\times 10^{1}$    \\
Model C4                    &
$10^{12}$                   & 1                       & $9.6\times 10^{42}$   &  $10^{0}$                   & $10^4$                &  
$4.8\times 10^{12}$         & $t_{\rm travel}$        & $t_{\rm travel}$      &  $5.9\times 10^{36}$        & $6.4\times 10^{1}$    \\
Model C5                    &
$10^{12}$                   & 1                       & $9.6\times 10^{42}$   &  $10^{0}$                   & $10^5$                &  
$4.8\times 10^{12}$         & $t_{\rm travel}$        & $t_{\rm travel}$      &  $6.0\times 10^{35}$        & $6.5\times 10^{1}$     \\
Model C6                    &
$10^{12}$                   & 1                       & $9.6\times 10^{42}$   &  $10^{0}$                   & $10^6$                &  
$4.8\times 10^{12}$         & $t_{\rm travel}$        & $t_{\rm travel}$      &  $6.0\times 10^{34}$        & $2.4\times 10^{1}$     \\
Model C7                    &
$10^{12}$                   & 1                       & $9.6\times 10^{42}$   &  $10^{0}$                   & $10^7$                &  
$4.8\times 10^{12}$         & $t_{\rm travel}$        & $t_{\rm travel}$      &  $6.0\times 10^{33}$        & $4.2\times 10^{0}$     \\
Model C8                    &
$10^{12}$                   & 1                       & $9.6\times 10^{42}$   &  $10^{0}$                   & $10^8$                &  
$4.8\times 10^{12}$         & $t_{\rm travel}$        & $t_{\rm travel}$      &  $6.0\times 10^{32}$        & $5.1\times 10^{-1}$    \\
Model C9                    &
$10^{12}$                   & 1                       & $9.6\times 10^{42}$   &  $10^{0}$                   & $10^9$                &  
$4.8\times 10^{12}$         & $t_{\rm travel}$        & $t_{\rm travel}$      &  $6.0\times 10^{31}$        & $5.1\times 10^{-2}$    \\
Model D1                    &
$10^{12}$                   & 1                       & $9.6\times 10^{42}$   &  $10^{-1}$                  & $10^1$                &  
$1.0\times 10^{6}$          & $t_{\rm col}$           & $t_{\rm ep}$          &  ------------               & ------------           \\
Model D2                    &
$10^{12}$                   & 1                       & $9.6\times 10^{42}$   &  $10^{-1}$                  & $10^2$                &  
$1.0\times 10^{6}$          & $t_{\rm col}$           & $t_{\rm ep}$          &  ------------               & ------------           \\
Model D3                    &
$10^{12}$                   & 1                       & $9.6\times 10^{42}$   &  $10^{-1}$                  & $10^3$                &  
$1.0\times 10^{6}$          & $t_{\rm col}$           & $t_{\rm col}$         &  ------------               & ------------            \\
Model D4                    &
$10^{12}$                   & 1                       & $9.6\times 10^{42}$   &  $10^{-1}$                  & $10^4$                &  
$1.0\times 10^{6}$          & $t_{\rm sync}$          & $t_{\rm sync}$        &  ------------               & ------------            \\
Model D5                    &
$10^{12}$                   & 1                       & $9.6\times 10^{42}$   &  $10^{-1}$                  & $10^5$                &  
$1.0\times 10^{6}$          & $t_{\rm sync}$          & $t_{\rm sync}$        &  ------------               & ------------            \\
Model D6                    &
$10^{12}$                   & 1                       & $9.6\times 10^{42}$   &  $10^{-1}$                  & $10^6$                &  
$1.0\times 10^{6}$          & $t_{\rm sync}$          & $t_{\rm sync}$        &  ------------               & ------------            \\
Model D7                    &
$10^{12}$                   & 1                       & $9.6\times 10^{42}$   &  $10^{-1}$                  & $10^7$                &  
$1.0\times 10^{6}$          & $t_{\rm sync}$          & $t_{\rm sync}$        &  ------------               & ------------            \\
Model D8                    &
$10^{12}$                   & 1                       & $9.6\times 10^{42}$   &  $10^{-1}$                  & $10^8$                &  
$1.0\times 10^{6}$          & $t_{\rm sync}$          & $t_{\rm sync}$        &  ------------               & ------------             \\
Model D9                    &
$10^{12}$                   & 1                       & $9.6\times 10^{42}$   &  $10^{-1}$                  & $10^9$                &  
$1.0\times 10^{6}$          & $t_{\rm sync}$          & $t_{\rm sync}$        &  ------------               & ------------           \\
Model E1                    &
$10^{15}$                   & 5                       & $1.5\times 10^{40}$   &  $10^{3}$                   & $10^{1}$                &  
$2.8\times 10^{13}$         & $t_{\rm travel}$        & $t_{\rm travel}$      &  $5.1\times 10^{33}$        & $3.0\times 10^{-9}$    \\
Model E2                    &
$10^{15}$                   & 5                       & $1.5\times 10^{40}$   &  $10^{3}$                   & $10^{2}$                &  
$2.8\times 10^{13}$         & $t_{\rm travel}$        & $t_{\rm travel}$      &  $6.7\times 10^{32}$        & $1.6\times 10^{-6}$    \\
Model E3                    &
$10^{15}$                   & 5                       & $1.5\times 10^{40}$   &  $10^{3}$                   & $10^{3}$                &  
$2.8\times 10^{13}$         & $t_{\rm travel}$        & $t_{\rm travel}$      &  $3.2\times 10^{31}$        & $1.1\times 10^{-5}$       \\
Model E4                    &
$10^{15}$                   & 5                       & $1.5\times 10^{40}$   &  $10^{3}$                   & $10^{4}$                &  
$2.8\times 10^{13}$         & $t_{\rm travel}$        & $t_{\rm travel}$      &  $3.2\times 10^{30}$        & $3.5\times 10^{-5}$    \\
Model E5                    &
$10^{15}$                   & 5                       & $1.5\times 10^{40}$   &  $10^{3}$                   & $10^{5}$                &  
$2.8\times 10^{13}$         & $t_{\rm travel}$        & $t_{\rm travel}$      &  $3.3\times 10^{29}$        & $3.5\times 10^{-5}$       \\
Model E6                    &
$10^{15}$                   & 5                       & $1.5\times 10^{40}$   &  $10^{3}$                   & $10^{6}$                &  
$2.8\times 10^{13}$         & $t_{\rm travel}$        & $t_{\rm travel}$      &  $3.3\times 10^{28}$        & $1.3\times 10^{-5}$     \\
Model E7                    &
$10^{15}$                   & 5                       & $1.5\times 10^{40}$   &  $10^{3}$                   & $10^{7}$                &  
$2.8\times 10^{13}$         & $t_{\rm travel}$        & $t_{\rm travel}$      &  $3.3\times 10^{27}$        & $2.3\times 10^{-6}$     \\
Model E8                    &
$10^{15}$                   & 5                       & $1.5\times 10^{40}$   &  $10^{3}$                   & $10^{8}$                &  
$2.8\times 10^{13}$         & $t_{\rm travel}$        & $t_{\rm travel}$      &  $3.3\times 10^{26}$        & $2.8\times 10^{-7}$    \\
Model F1                    &
$10^{12}$                   & 5                       & $1.5\times 10^{40}$   &  $10^{2}$                   & $10^1$                &  
$8.9\times 10^{13}$         & $t_{\rm travel}$        & $t_{\rm travel}$      &  $1.4\times 10^{33}$        & $8.3\times 10^{-10}$   \\
Model F2                    &
$10^{12}$                   & 5                       & $1.5\times 10^{40}$   &  $10^{2}$                   & $10^2$                &  
$8.9\times 10^{13}$         & $t_{\rm travel}$        & $t_{\rm travel}$      &  $1.8\times 10^{32}$        & $4.2\times 10^{-7}$     \\
Model F3                    &
$10^{12}$                   & 5                       & $1.5\times 10^{40}$   &  $10^{2}$                   & $10^3$                &  
$8.9\times 10^{13}$         & $t_{\rm travel}$        & $t_{\rm travel}$      &  $8.7\times 10^{30}$        & $2.9\times 10^{-6}$    \\
Model F4                    &
$10^{12}$                   & 5                       & $t_{\rm travel}$      &  $10^{2}$                   & $10^4$                &  
$8.9\times 10^{13}$         & $t_{\rm travel}$        & $t_{\rm travel}$      &  $8.8\times 10^{29}$        & $9.4\times 10^{-6}$    \\
Model F5                    &
$10^{12}$                   & 5                       & $1.5\times 10^{40}$   &  $10^{2}$                   & $10^5$                &  
$8.9\times 10^{13}$         & $t_{\rm travel}$        & $t_{\rm travel}$      &  $8.8\times 10^{28}$        & $9.5\times 10^{-6}$     \\
Model F6                    &
$10^{12}$                   & 5                       & $1.5\times 10^{40}$   &  $10^{2}$                   & $10^6$                &  
$8.9\times 10^{13}$         & $t_{\rm travel}$        & $t_{\rm travel}$      &  $8.8\times 10^{27}$        & $3.6\times 10^{-6}$     \\
Model F7                    &
$10^{12}$                   & 5                       & $1.5\times 10^{40}$   &  $10^{2}$                   & $10^7$                &  
$8.9\times 10^{13}$         & $t_{\rm travel}$        & $t_{\rm travel}$      &  $8.8\times 10^{26}$        & $6.2\times 10^{-7}$    \\
Model F8                    &
$10^{12}$                   & 5                       & $1.5\times 10^{40}$   &  $10^{2}$                   & $10^8$                &  
$8.9\times 10^{13}$         & $t_{\rm travel}$        & $t_{\rm travel}$      &  $8.8\times 10^{25}$        & $7.6\times 10^{-8}$     \\
Model G1                    &
$10^{12}$                   & 5                       & $1.5\times 10^{40}$   &  $10^{1}$                   & $10^1$                &  
$8.9\times 10^{13}$         & $t_{\rm travel}$        & $t_{\rm travel}$      &  $3.4\times 10^{32}$        & $2.1\times 10^{-10}$   \\
Model G2                    &
$10^{12}$                   & 5                       & $1.5\times 10^{40}$   &  $10^{1}$                   & $10^2$                &  
$3.0\times 10^{14}$         & $t_{\rm travel}$        & $t_{\rm travel}$      &  $4.3\times 10^{31}$        & $1.0\times 10^{-7}$    \\
Model G3                    &
$10^{12}$                   & 5                       & $1.5\times 10^{40}$   &  $10^{1}$                   & $10^3$                &  
$3.0\times 10^{14}$         & $t_{\rm travel}$        & $t_{\rm travel}$      &  $2.1\times 10^{30}$        & $7.2\times 10^{-7}$    \\
Model G4                    &
$10^{12}$                   & 5                       & $1.5\times 10^{40}$   &  $10^{1}$                   & $10^4$                &  
$3.0\times 10^{14}$         & $t_{\rm travel}$        & $t_{\rm travel}$      &  $2.1\times 10^{29}$        & $2.3\times 10^{-6}$    \\
Model G5                    &
$10^{12}$                   & 5                       & $1.5\times 10^{40}$   &  $10^{1}$                   & $10^5$                &  
$3.0\times 10^{14}$         & $t_{\rm travel}$        & $t_{\rm travel}$      &  $2.1\times 10^{28}$        & $2.3\times 10^{-6}$    \\
Model G6                    &
$10^{12}$                   & 5                       & $1.5\times 10^{40}$   &  $10^{1}$                   & $10^6$                &  
$3.0\times 10^{14}$         & $t_{\rm travel}$        & $t_{\rm travel}$      &  $2.1\times 10^{27}$        & $8.8\times 10^{-7}$    \\
Model G7                    &
$10^{12}$                   & 5                       & $1.5\times 10^{40}$   &  $10^{1}$                   & $10^7$                &  
$3.0\times 10^{14}$         & $t_{\rm travel}$        & $t_{\rm travel}$      &  $2.1\times 10^{26}$        & $1.5\times 10^{-7}$    \\
Model G8                    &
$10^{12}$                   & 5                       & $1.5\times 10^{40}$   &  $10^{1}$                   & $10^8$                &  
$3.0\times 10^{14}$         & $t_{\rm travel}$        & $t_{\rm travel}$      &  $2.1\times 10^{25}$        & $1.8\times 10^{-8}$     \\
Model H1                    &
$10^{12}$                   & 5                       & $1.5\times 10^{40}$   &  $10^{0}$                   & $10^1$                &  
$1.0\times 10^{6}$          & $t_{\rm travel}$        & $t_{\rm ep}$          &  ------------               & ------------          \\
Model H2                    &
$10^{12}$                   & 5                       & $1.5\times 10^{40}$   &  $10^{0}$                   & $10^2$                &  
$1.0\times 10^{6}$          & $t_{\rm travel}$        & $t_{\rm ep}$          &  ------------               & ------------          \\
Model H3                    &
$10^{12}$                   & 5                       & $1.5\times 10^{40}$   &  $10^{0}$                   & $10^3$                &  
$1.0\times 10^{6}$          & $t_{\rm travel}$        & $t_{\rm travel}$      &  $1.5\times 10^{39}$        & $4.9\times 10^{2}$     \\
Model H4                    &
$10^{12}$                   & 5                       & $1.5\times 10^{40}$   &  $10^{0}$                   & $10^4$                &  
$1.0\times 10^{6}$          & $t_{\rm travel}$        & $t_{\rm travel}$      &  $1.5\times 10^{38}$        & $1.6\times 10^{3}$     \\
Model H5                    &
$10^{12}$                   & 5                       & $1.5\times 10^{40}$   &  $10^{0}$                   & $10^5$                &  
$1.0\times 10^{6}$          & $t_{\rm sync}$          & $t_{\rm sync}$        &  ------------               & ------------          \\
Model H6                    &
$10^{12}$                   & 5                       & $1.5\times 10^{40}$   &  $10^{0}$                   & $10^6$                &  
$1.0\times 10^{6}$          & $t_{\rm sync}$          & $t_{\rm sync}$        &  ------------               & ------------           \\
Model H7                    &
$10^{12}$                   & 5                       & $1.5\times 10^{40}$   &  $10^{0}$                   & $10^7$                &  
$1.0\times 10^{6}$          & $t_{\rm sync}$          & $t_{\rm sync}$        &  ------------               & ------------           \\
Model H8                    &
$10^{12}$                   & 5                       & $1.5\times 10^{40}$   &  $10^{0}$                   & $10^8$                &  
$1.0\times 10^{6}$          & $t_{\rm sync}$          & $t_{\rm sync}$        &  ------------               & ------------           \\
\enddata
\tablenotetext{a}{
Shortest timescale among proton's synchrotron cooling timescale $t_{\rm sync}$, traveling timescale $t_{\rm travel}$, and collision timescale
$t_{\rm col}$. When $t_{\rm min}$ = $t_{\rm travel}$ or $t_{\rm col}$, it means that $t_{\rm min}$ is always shorter than $t_{\rm sync}$.
On the other hand, when $t_{\rm min}$ = $t_{\rm col}$, it means that $t_{\rm col}$ is shorter than any other timescales at some region of the
nebula flow.
}
\tablenotetext{b}{
Shorter timescale between $t_{\rm min}$ and energy transfer timescale from protons to electrons $t_{\rm ep}$.
When min($t_{\rm min}$,$t_{\rm ep}$) = $t_{\rm min}$, it means that $t_{\rm min}$ is always shorter than $t_{\rm ep}$
in the nebula flow. On the other hand, when  min($t_{\rm min}$,$t_{\rm ep}$) = $t_{\rm ep}$, it means that $t_{\rm ep}$
is shorter than $t_{\rm min}$ at some region of the nebula flow.
}
\tablecomments{
Luminosity of pions ($L_{\pi}$) and event rate at a km$^3$ high-energy neutrino detector are not shown when the conditions (i)-(iii)
presented in section 3 are not satisfied.
}
\label{tab2}
\end{deluxetable}

\end{document}